\documentclass{aa}
\usepackage[varg]{txfonts}
\usepackage{natbib}
\usepackage{longtable}
\bibpunct{(}{)}{;}{a}{}{,}
\usepackage{epstopdf}
\usepackage{graphicx}
\usepackage{subfig}
\usepackage{array}

\begin{document}
\title{Variable magnetic field geometry of the young sun HN Peg (HD 206860)}
\author{S.~Boro Saikia\inst{\ref{inst1}}\and S.~V.~Jeffers\inst{\ref{inst1}}\and
P.~Petit\inst{\ref{inst2}}\and S.~Marsden\inst{\ref{inst3}} \and J.~Morin\inst{\ref{inst1},\ref{inst4}} \and C.~P.~Folsom\inst{\ref{inst2}}}

\institute{Institut f\"ur Astrophysik, Universit\"at G\"ottingen, Friedrich Hund Platz 1, 37077 G\"ottingen, Germany\label{inst1}
\and CNRS, Institut de Recherche en Astrophysique et Plan\'etologie, 14 Avenue Edouard Belin, F-31400 Toulouse, France \label{inst2}
\and Computational Engineering and Science Research Centre, University of Southern Queensland, Toowoomba 4350, Australia \label{inst3}
\and LUPM-UMR 5299,CNRS \& Universit\'e Montpellier, Place Eug\'ene Bataillon, 34095 Montpellier Cedex 05, France \label{inst4}
}
\date{}

\abstract{The large-scale magnetic field of solar-type stars reconstructed from their spectropolarimetric observations provide important insight into their underlying dynamo processes.}
{We aim to investigate the temporal variability of the large-scale surface magnetic field and chromospheric activity of a young solar analogue, the G0 dwarf HN Peg.}
{The large-scale surface magnetic field topology is reconstructed using Zeeman Doppler Imaging at six observational epochs covering seven years.
We also investigated the chromospheric activity variations by measuring the flux
in the line cores of the three chromospheric activity indicators: Ca II H$\&$K, H$\alpha$, and the Ca II IRT lines. }
{ The magnetic topology of HN Peg shows a complex and variable geometry. While
the radial field exhibits a stable positive polarity magnetic region at the poles at each observational epoch, the azimuthal field is strongly variable in strength, where a strong band of positive polarity magnetic field
is present at equatorial latitudes. This field disappears during the middle of our timespan, reappearing again during the last two epochs of observations. 
The mean magnetic field derived from the magnetic maps also follow a similar trend to the toroidal field, with the field strength at a minimum in epoch 2009.54. Summing the line of sight magnetic field over the visible
surface at each observation, HN Peg exhibits a weak longitudinal magnetic field (B$_l$) ranging from -14 G to 13 G, with no significant long-term trend, although there is significant rotational
variability within each epoch. Those chromospheric activity indicators exhibit more long-term variations over the time span of observations, where the minimal is observed in Epoch 2008.71.
  }{}

\maketitle

\section{Introduction}

Solar dynamo models suggest that the regeneration of the solar magnetic field results from the interplay between convection and differential rotation \citep{parker,brandenburg,charbonneau}. These cyclic dynamo processes
are responsible for the different manifestations of solar activity, such as prominences, flares, and solar winds. 
Observations of the surface activity features as above of young solar-type stars provide an important insight into the underlying dynamo processes that operate in stars other than the Sun \citep [see][]{donatilandstreet}.
In general, solar-type stars all have a similar internal structure to the Sun, with a radiative core surrounded by a convective envelope. This would suggest that, as for the Sun, their magnetic activity is 
generated by  a $\mathrm{\alpha}$-$\mathrm{\Omega}$ dynamo. Although the exact mechanism of the dynamo processes is still not completely understood, observations of significant azimuthal field in rapidly rotating solar-type
stars indicate the presence of dynamo distributed throughout the convective zone \citep{donati2003,pascal08}. The possibility that such a distributed dynamo operates in rapidly rotating solar-type stars is supported by 
detailed numerical modelling \citep{brown10}.

\paragraph{}
The presence of the magnetic field can result in emission in the line cores of certain chromospheric lines, such as Ca II H$\&$K, H$\alpha$, and Ca II IRT lines. The Mount Wilson survey was the first long-term monitoring of
Ca II H$\&$K to investigate the surface magnetic variability of stars with an outer convective envelope \citep{baliunas1995,duncan90}. This survey observed a wide range of variability ranging from cyclic and irregular
variability to no variations at all. Since the Mount Wilson project, a host of other activity surveys have been carried out in the past few decades \citep{wright04,hall07}. 
While monitoring Ca II H$\&$K can provide a long-term indication of the magnetic variability, the disadvantage of using this method is that this tracer
does not provide any direct measurement of the field strength or information about the large-scale geometry of the star's magnetic field. 
\paragraph{}
In recent years with the arrival of spectropolarimeters, such as ESPaDOnS, NARVAL, and HARPSpol, the magnetic field observations of other solar-type stars have helped in understanding the dynamo processes that drive
the different manifestations of surface
activity \citep{pascal08,marsden}. For example, magnetic cycles have been observed on the F7 dwarf $\tau$ Bootis (1.42 $\pm$ 0.05 M$_\sun$, $\it {T}_\mathrm{eff}$=6360 $\pm$ 80 K \citep{catala}), 
where the large-scale magnetic field is observed to exhibit cycles over a two-year period \citep{fares}. Magnetic cycles have also been observed on HD 78366 \citep{morgenthaler11} with a mass of 1.34 $\pm$ 0.13 M$_\sun$ and
$\it {T}_\mathrm{eff}$ of 6014$\pm$50 K, which shows two polarity reversals  with a probable cycle of approximately 
three years.  More complex variability has been observed on HD 190771 \citep{pascal08,petit09}, which is similar in mass to the Sun (0.96$\pm$0.13 M$_\sun$) with $\it {T}_\mathrm{eff}$ of 5834$\pm$50 K, where polarity reversals
are observed in its radial and azimuthal field, but 
over the time span of the observations, they do not return to the initial field configuration.  These results indicate that the 22-year magnetic cycle of the Sun is not an exception but that cyclic activity is also 
present in other solar type stars with ages close to the Sun's.
\paragraph{}
In this paper we determine the large-scale magnetic field geometry of the young solar analogue HN Peg using the technique of Zeeman-Doppler Imaging.    
We also measure HN Peg's chromospheric activity using the core emission in Calcium II H$\&$K, H$\alpha$, and Calcium II IR triplet lines.  In Section 2 we review the literature on HN Peg, followed 
by a description of the observations in Section 3 and the longitudinal magnetic field in Section 4. The chromospheric activity measurements are described and presented in Section 5, and the large-scale magnetic field reconstructions are 
presented in Section 6.  The results are discussed in Section 7.

\section{HN Peg}
HN Peg is a G0 dwarf with a mass of 1.085 $\pm$ 0.091 M$_\sun$ and a radius of 1.002 $\pm$ 0.018 R$_\sun$ \citep{fischer}, as shown in Table~\ref{table:1}. It is part of the Her-Lyr association. 
HN Peg's association with the Her-Lyr moving group was discovered by \citet{gaidos} when he detected a group of stars 
(V439 And, MN UMa, DX Leo, NQ UMa, and HN Peg) with similar kinematics \citep{fuhrmann,lopez}. The age of Her-Lyr was established to be approximately 200 Myr by gyrochronology and also by comparing the Li and 
H$\alpha$ lines of Her-Lyr with the UMa group \citep{fuhrmann,eisen13}. A separate gyrochronology study carried out on the Mount Wilson sample also 
provided an age of HN Peg of 237$\pm$33 Myr  \citep{barnes07}.
 
\paragraph{}
The Mount Wilson survey estimated a period of 6.2$\pm$0.2 years for HN Peg, with high chromospheric variability \citep{baliunas1995,schroeder}. Photometric measurements 
carried out by \citet{photometry} claimed there is a solar-type star spot cycle of HN Peg with a period of 5.5 $\pm$ 0.3 years. Both spectroscopic and photometric observations of HN Peg  were observed by \citet{frasca}, where rotational modulation 
in both Ca II H$\&$K and H$\alpha$ were observed. Power spectrum analysis of the spectra of HN Peg \citep{baliunas85} suggests the presence of surface differential rotation. Differential rotation of HN Peg was 
also investigated by observing variations in the rotational period \citep{antisolar}, where the evolution of the average rotation of HN Peg along the activity cycle was observed to be anti-solar.
\paragraph{}
From direct imaging using the Spitzer Space Telescope, HN Peg also has also been shown to have an early T-dwarf companion HN Peg b 
at  a distance of approximately 794 AU \citep{luhman,browndwarf}. Photometric observations of HN Peg have indicated that it also harbours a debris disk with a steep spectral energy distribution \citep{ertel}.

\section{Observations}
The data were collected as part of the international Bcool collaboration \footnote{\url{http://bcool.ast.obs-mip.fr/Bcool}} \citep{marsden}, using the NARVAL spectropolarimeter at the 2 m Telescope Bernard Lyot (TBL) at Pic du Midi Observatory. 
NARVAL is a new generation spectropolarimeter which is a twin of the ESPaDOnS stellar spectropolarimeter. NARVAL is a cross dispersed
echelle spectrograph with minimum instrumental polarisation \citep{auriere2003,pascal08}. NARVAL has a resolution of approximately 65\,000 and covers the 
full optical domain from 370 nm to 1000 nm, ranging over 40 grating orders.

\paragraph{}
The data were extracted using Libre-ESpRIT \citep{donati1997}, which is a fully automated data reduction package installed at TBL. The observations were taken to maximise rotational phase coverage. 
Seven sets of data were obtained for the observational epochs 2007.67, 2008.71, 2009.54, 2010.62, 2011.67, 2012.61 and 2013.68, Table~\ref{kstars}. Each of these seven epochs contains 8 to 14 polarised Stokes V
observations.
\paragraph{}
Because the S/N in individual spectral lines is not high enough to detect Zeeman signatures in polarised light, we apply the technique of least square deconvolution (LSD) on the spectra \citep{donati1997,kochukhov10}. 
LSD is a multi-line technique which considers a similar local profile for each line  and obtains an averaged line profile by deconvolving the stellar spectra to a line mask.
A G2 line mask consisting of approximately 4800 lines matching a stellar photosphere model for HN Peg was used to generate the averaged line profile for Stokes I and Stokes V, resulting in huge multiplex gain in the S/N ratio of 
the polarised Stokes V profile as shown in Table~\ref{kstars}. The mask covers a wavelength range of 370 nm to 900 nm and the LSD profiles are normalised by using a mean Land\'e factor of 1.21 and a mean wavelength of 
550 nm from the line list.

\paragraph{}
The polarised Stokes V spectra from 2012 and part
of 2011 were discarded because of instrumental defects of NARVAL because the reference point for one of the polarisation rhombs was incorrect.
This resulted in incorrect polarisation signatures for HN Peg in the last two observations in 2011 (Table~\ref{kstars}) and sudden polarisation changes in 2012. However the unpolarised spectra in 2011 and 2012 are not affected,
and can be used to measure the chromospheric proxies of magnetic activity. Subsequent tests have confirmed that the polarised data collected in 2013 is reliable.\footnote{A detailed description of the polarisation 
defects and the correction technique used can be found here: \url{http://spiptbl.bagn.obs-mip.fr/Actualites/Anomalies-de-mesures}} 

  \begin{table}
    \caption{Summary of the physical parameters of HN Peg.}              
\label{table:1}  
\begin{tabular}{l l c }          
\hline\hline                        
Parameters &  HD 206860 & References \\    
\hline                                   
    Effective temperature, {T$_\text{eff}$}& 5974 (K) & 1 \\
    Radius & 1.002$\pm$0.018 (R$_\sun$)& 1   \\
    Mass & 1.085$\pm$0.091 (M$_\sun$)& 1 \\
    Rotational period, { P}  & 4.6 (days)& this work\\
    Rotational velocity, { v$\sin$ i} & 10.6 (kms$^{-1}$)& 1\\
    Age & $\sim$200 (Myr)& 2\\
     \hline                                             
\end{tabular}
\tablebib{
(1)~\citet{fischer};(2)~\citet{age} and \citet{eisen13}.
}
\end{table}

\section{Mean longitudinal magnetic field (B$_{l}$)}
\begin{figure}
\centering
\includegraphics[width = .55\textwidth]{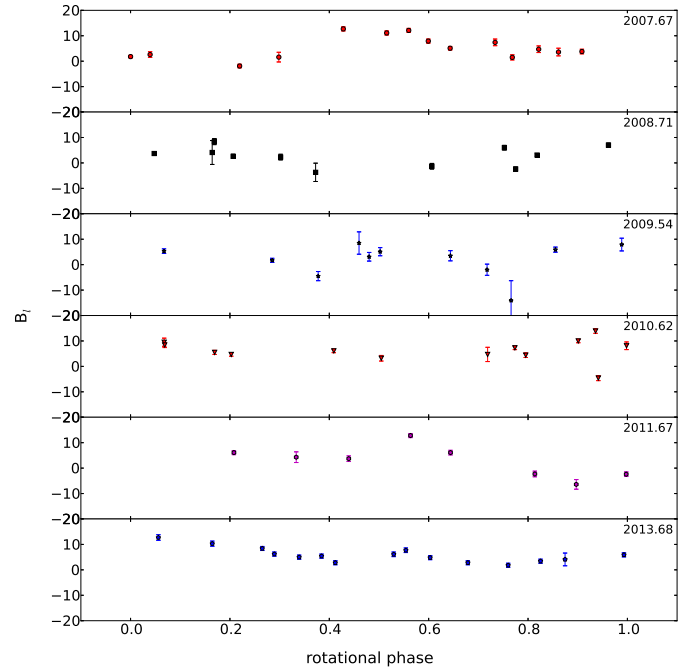}
  \caption{Variability of the longitudinal field ($B_l$) as a function of the rotational phase. Each of 
  the sub plots from top to bottom correspond to six different epochs (2007.67, 2008.71, 2009.54, 2010.62, 2011.67, and 2013.68). } 
\label{fields}
\end{figure}

The longitudinal magnetic field is measured using the LSD Stokes V and Stokes I profiles, where the field measured is the mean magnetic field (line of sight component) integrated over
the entire visible stellar surface.  
The centre-of-gravity method \citep{resemel} was used on the LSD profile of HN Peg to measure its longitudinal magnetic field, as shown in equation~\ref{magnetic},
 \begin{equation}
  \mathrm{B}_l(\mathrm{G}) = -2.14\times 10^{11}\frac{\int v V(v) dv}{\mathrm \lambda_0 \mathrm g \mathrm c \int (I_c - I(v)) dv}
  \label{magnetic}
 \end{equation}
where B$_{l}$ is the longitudinal magnetic field in Gauss, $\lambda_0$ = 550 nm is the central wavelength of the LSD profile, $\it g$ = 1.21 is the Land\'e factor of the line list, $\it c$ is the speed of light in kms$^{-1}$,
 $\it v$  is the radial velocity in kms$^{-1}$ and $\it I_c$ is the normalised continuum. The velocity range covered by the integration window is $\pm 17$ kms$^{-1}$ around the line centre.
 The uncertainty in each of the longitudinal magnetic field measurements are obtained by propagating the errors computed by the reduction pipeline in equation~\ref{magnetic} as described by \citet{marsden}.
 The magnetic field from the LSD null profiles were also calculated for each observations where the magnetic field is approximately zero, indicating negligible spurious polarisation affect on the longitudinal field measurements. 
 The errors in the longitudinal magnetic field of HN Peg is higher than the null profiles except in a few cases where the SN is weak compared to the rest of the observations. 
 The magnetic field of HN Peg (B$_{l}$) and the magnetic field from the null profile and their related 
 uncertainties are recorded in Table~\ref{values}. 
\paragraph{}
 The variability of the longitudinal magnetic field  of HN Peg as a function of rotational phase is shown in Fig~\ref{fields}. The phase dependence of the longitudinal magnetic field indicates a complex surface magnetic field geometry. 
No long-term trend in mean longitudinal field measurements was observed for HN Peg as shown in Fig~\ref{mean}, where B$_{l}$ exhibits no significant long-term variations through out the observational timespan. 
The mean B$_{l}$ value of 5.3 G with a dispersion of 4.2 G in epoch 2007.67 goes down to its lowest value of 1.9 G with a dispersion of 6.3 G in epoch 2009.54. The mean value is the highest in epoch 2010.62. 
 
\section{Chromospheric activity indicators}

Chromospheric activity has been widely observed in solar type stars, which is manifested as emission in the line cores of the chromospheric lines, such as: Ca II H$\&$K, H$\alpha$, and Ca II IRT lines. 
The  varying flux in these line cores can be used as a proxy to investigate magnetic cycles. 

\subsection{S-index}
We observe strong emission lines in the Ca II H$\&$K line cores of HN Peg  as a function of its rotational phase. The S-index is calculated by using two triangular band passes centred at Ca II H and K lines \citep{duncan90,morgenthaler12}
at 396.8469 and 393.3663 nm respectively with a FWHM of 0.1 nm. The flux in the continuum  at the red and blue sides of the line is measured by using two 2 nm wide rectangular band passes R and V at 400.107 and 390.107 nm respectively. Equation~\ref{sindex} 
is used to calibrate our index with the Mount Wilson  values,
\begin{equation}
 \mathrm{\text{S-index}}=\frac {\alpha \mathrm{F_H} + \beta \mathrm{F_K}} {\gamma \mathrm{F_R} + \delta \mathrm{F_V}} + \varPhi
 \label{sindex}
\end{equation}
where F$_\mathrm{H}$, F$_\mathrm{K}$, F$_\mathrm{R}$ and F$_\mathrm{V}$ are the flux in the band passes H, K, R and V. The NARVAL coefficients used to match our S-index values to the Mt Wilson values \citep{marsden}
are  $\alpha$ = 12.873, $\beta$ = 2.502, $\gamma$ = 8.877, $\delta$= 4.271 and $\varPhi$ = 1.183e-03.
We do not carry out the renormalisation procedure used by \citet{morgenthaler12} and carry out the continumm check following the procedure in \citet{waite}, where it was determined that removal of the 
overlapping orders is as efficient as renormalisation of the spectra.

 \begin{figure}
 \centering
 \includegraphics[width = .55\textwidth]{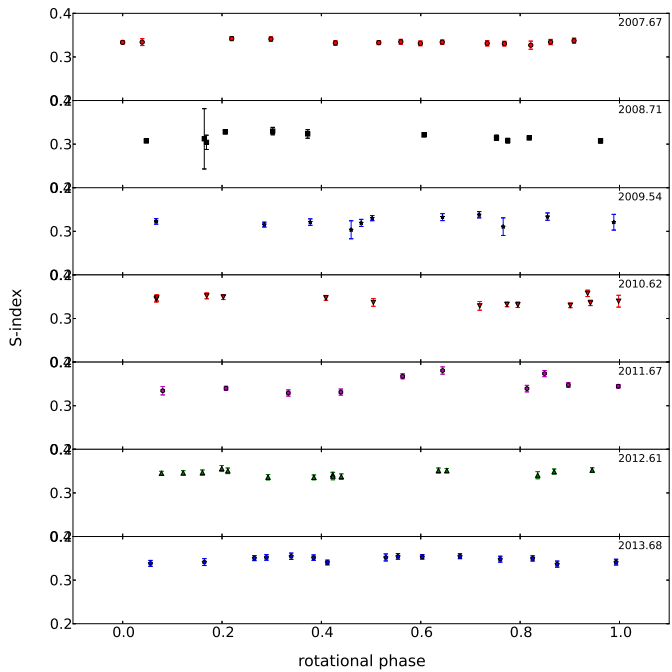}
\caption{Variability of S-index as a function of the rotational phase. Each of the sub plots from top to bottom correspond to seven different 
epochs (2007.67, 2008.71, 2009.54, 2010.62, 2011.67, 2012.61, and 2013.68).} 
 \label{varsindex}
 \end{figure}
 
\paragraph{}
The variability of HN Peg's S-index for each of the seven epochs is shown as a function of rotational phase  in Fig~\ref{varsindex}. The error in the S-index for each measurement was calculated using error propagation.
The S-index and related uncertainty for each observation is shown in Table~\ref{values}.

\begin{figure}
  \centering
 \includegraphics[width=.55\textwidth]{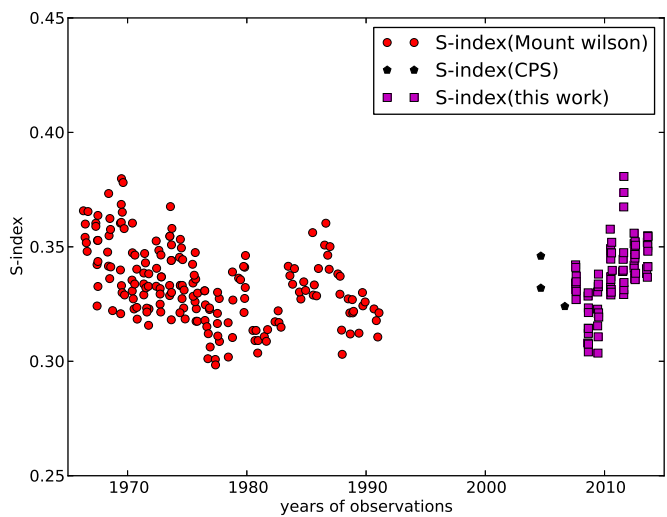}
 \caption{S-index measurements of HN Peg from the combined data sets. The red circles represent data from the Mount Wilson survey, the black hexagons represent data from the CPS survey and the magenta squares are our measurements.}
 \label{totalsindex}
\end{figure}
\paragraph{}
We also included S-index measurements of HN Peg from the Mount Wilson survey, where the data was collected from 1966 to 1991 \citep{baliunas1995}. There are no published S-index measurements
of HN Peg from 1991 to 2004. Additional S-index values were obtained from the California planet search
program \citet{isaacson}, two in 2004 and one in 2006. No error bars are available for the S-index measurements taken from literature.
The  long term S-index measurements from the combined data are shown in Fig~\ref{totalsindex}.  

\subsection{H$\alpha$-index}

\begin{figure}
 \centering
 \includegraphics[width = .55\textwidth]{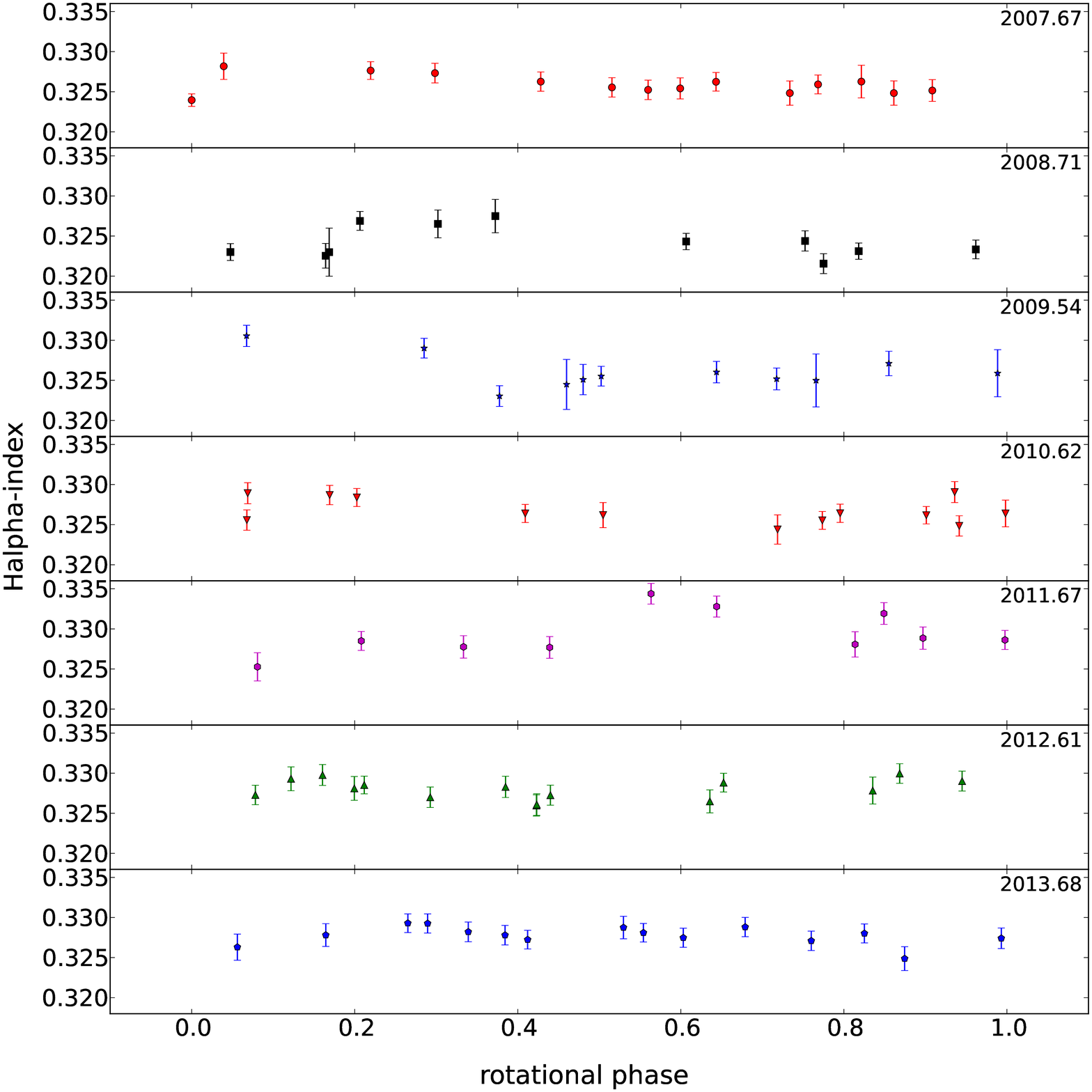}
 \caption{Variability of the H$\mathrm{\alpha}$-index as a function of rotational phase. Each of the sub plots from top to bottom
 correspond to seven different epochs (2007.67, 2008.71, 2009.54, 2010.62, 2011.67, 2012.61, and 2013.68).}
 \label{ha}
 \end{figure}

The rotational variation of the H$\alpha$ line was also determined.
 A rectangular band pass of 0.36 nm width, centred at the H$\alpha$ line at 656.285 nm \citep{gizis02} and two 0.22 nm wide rectangular band passes H$_{blue}$ and H$_{red}$ at 655.77 and 656.0 nm 
 respectively were used to measure the H${\alpha}$-index. We corrected the order overlap in the NARVAL spectra and used the order containing the complete H$\alpha$ line core. We then calculated the H$_{\alpha}$-index 
 using equation \ref{halpha},
\begin{equation}
  \mathrm{H\alpha\text{-index}} = \frac {\mathrm{F_{H\alpha}}} {\mathrm{F_{blue}+F_{red}}}
\label{halpha}
 \end{equation}
 where F$_\mathrm{H\alpha}$ represents the flux in the H$\alpha$ line core and F$_{blue}$ and F$_{red}$  represent the flux in the continuum band pass filters  H$_{blue}$ and H$_{red}$ 
 respectively. The variability of H$\alpha$ as a function of HN Peg's rotational phase is shown in Fig~\ref{ha}. The uncertainty in H$\alpha$-index measurements were calculated using error propagation. The H$\alpha$-index 
 and related uncertainty for each observations are shown in Table~\ref{values}.

\subsection{CaIRT-index}
Since NARVAL covers a wide wavelength range from 350 nm upto 1000 nm, we can also observe the Ca II IR triplet lines. We take 0.2 nm wide rectangular band passes in the cores of each of the triplet lines at 849.8023 nm, 854.2091 nm 
and 866.241 nm. Two continuum band passes of the width of 0.5 nm are defined as IR$_{red}$ at 870.49 nm and IR$_{blue}$ at 847.58 nm for the flux at the red and blue sides of the 
IR lines \citep{2013LNP}. We calculate the Ca$\mathrm{IRT}$-index using equation~\ref{ir},
\begin{equation}
\mathrm {\text{CaIRT-index}} = \frac {\mathrm{F_{Ca1}+F_{Ca2}+F_{Ca3}}}{\mathrm{F_{IRT_{blue}}+F_{IRT_{red}}}}
 \label{ir}
\end{equation}

\begin{figure}
 \centering
 \includegraphics[width = .55\textwidth]{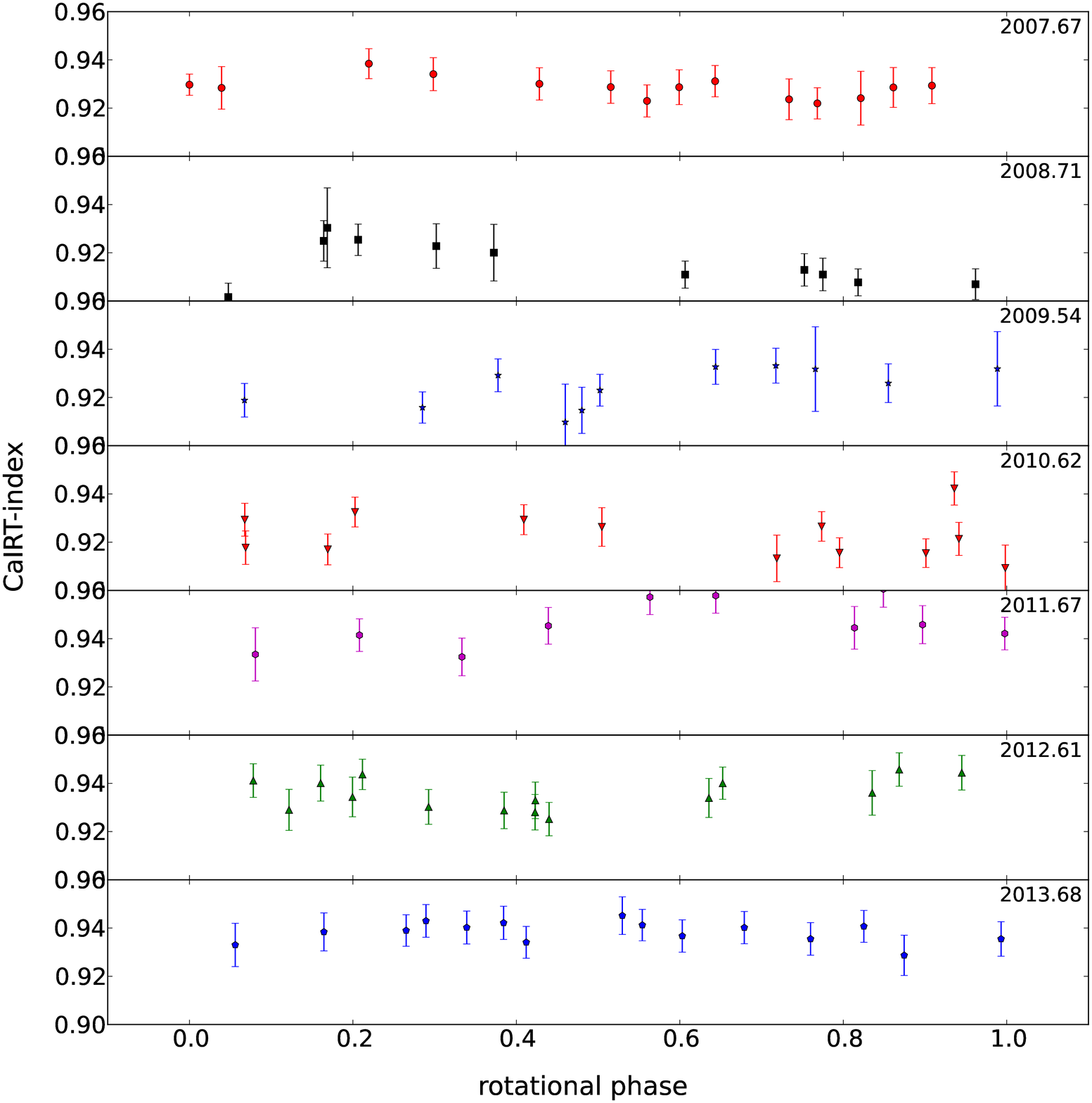}
  \caption{Variability of Ca$\mathrm{IRT}$-index  with the rotational phase. Each of the sub plots from top to bottom correspond to seven different epochs (2007.67, 2008.71, 2009.54, 2010.62, 2011.67, 2012.61, and 2013.68).} 
\label{irt}
 \end{figure}

where the flux in the three line cores are represented by F$_\mathrm{Ca1}$, F$_\mathrm{Ca2}$ and F$_\mathrm{Ca3}$ respectively and the continuum fluxes are defined by F$_\mathrm{IRT_{blue}}$ and 
F$_\mathrm{IRT_{red}}$ respectively. The error bars for individual observations were calculated using error propagation . 
The variability of the Ca$\mathrm{{IRT}}$-index as a function of HN Peg's rotational phase is shown in Fig ~\ref{irt}.
 
\begin{figure}
 \centering
 \includegraphics[width = .55\textwidth]{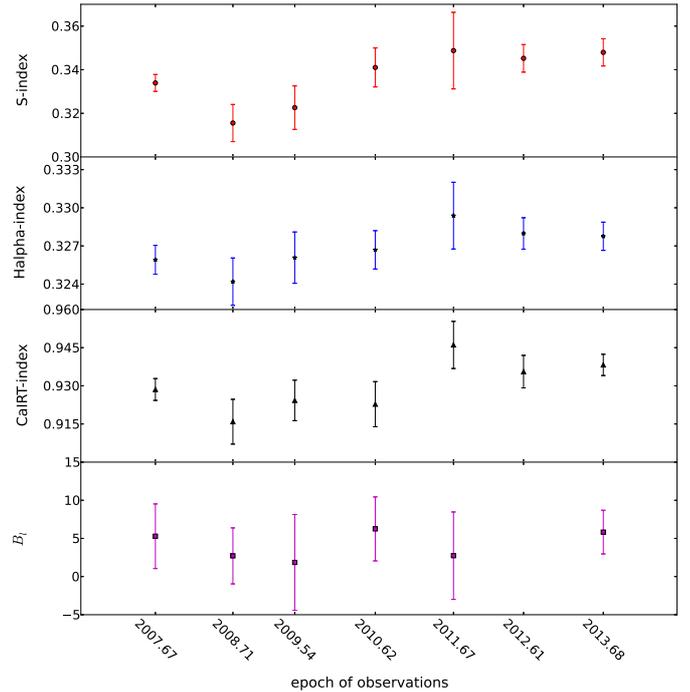}
\caption{Average value of the three different indices with the vertical bars showing the dispersion in each epoch of observations. Top to Bottom: Average values of S-index(red full circles),H$\alpha$-index
(blue stars), CaIRT-index(black triangle) and B$_l$(magenta squares) plotted against the epochs(2007.67-2013.68) in average Julian dates.}
\label{mean}
 \end{figure}

\begin{figure}
  \centering
  \includegraphics[width=.55\textwidth]{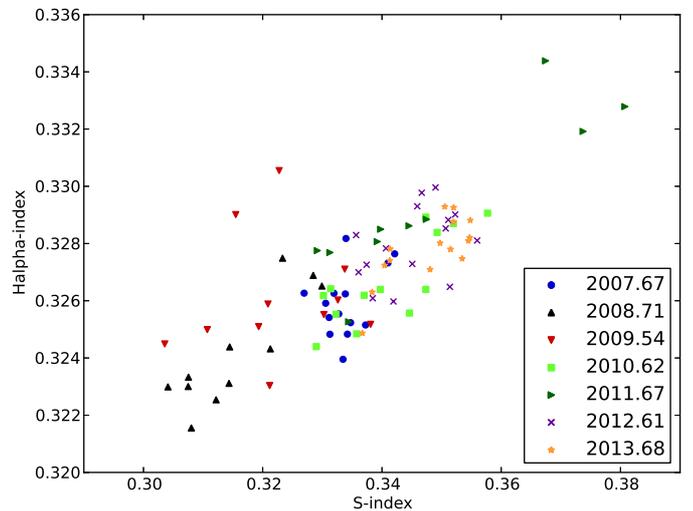}
  \caption{Correlation between the S-index and H$\alpha$-index for each epoch of observations.}
  \label{sha}
 \end{figure}

\begin{figure}
  \centering
  \includegraphics[width=.55\textwidth]{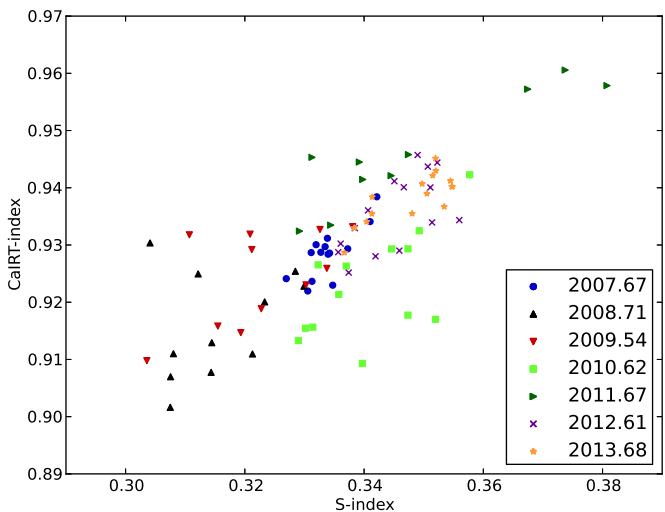} 
  \caption{Correlation between the S-index and CaIRT-index for each epoch of observations.}
  \label{sirt}
 \end{figure}

\paragraph{}
The long-term variability over the observational epochs of this analysis for the three activity indicators: Ca II H$\&$K, H$\alpha$, and Ca II IRT lines are shown in Fig~\ref{mean}, where the mean values of each index is plotted 
as a function of the observational epochs. The error bars represent the standard deviation of the activity proxies for each epoch of observation. The long-term S-index variations are prominent than the rotationally induced variations. 
The long-term S-index and H$\alpha$-index show visible correlation over the entire time span of our observations. The two indices show a decreasing trend from 2007.67 to 2008.71 and show an
increasing trend from 2008.71 to 2011.67 and then exhibits a flat trend. The long-term CaIRT-index measurements show global correlation with the S-index but shows more small-scale variations
on a year-to-year basis.
 \paragraph{}
 The correlation between the different activity proxies for individual epochs is prominent in some epochs but not clearly visible in other epochs. The correlation between S-index and H$\alpha$-index throughout the entire 
observational timespan is shown in Fig~\ref{sha}, where the correlation is more clearly visible. Correlation between S-index and CaIRT-index is shown in Fig~\ref{sirt}. The activity index measurements and their related uncertainties of  HN Peg is tabulated in Table~\ref{values}.

\section{Large-scale magnetic field topology}
The large-scale surface magnetic topology of HN Peg was reconstructed by using the Zeeman Doppler imaging (ZDI) tomographic technique, developed by \citet{semel89,brown91,zdi1997}. ZDI technique involves solving an inverse
problem and reconstructing the large-scale surface magnetic geometry by  iteratively comparing the Stokes V profile to the  synthetically generated profiles, which are generated from a synthetic stellar model. 
\paragraph{}
The large-scale field geometry of HN Peg was reconstructed by using the version of ZDI that reconstructs the field into its toroidal and poloidal components, expressed as spherical harmonics expansion \citep{donati06}.
A synthetic stellar model of HN Peg was constructed using 5000 grid points, where the local Stokes I profile in each grid cell was assumed to have a Gaussian shape and was adjusted to match the observed Stokes I profile. 
The synthetic local Stokes V profiles were computed under the weak field assumption and iteratively compared to the observed Stokes V profile.
The maximum entropy approach adopted by the ZDI code is based on the algorithm of \citet{skilling}. In this implementation of the maximum entropy principle, a target value of the reduced $\chi^2$ is set by the
user, where we define the reduced $\chi^2$ as the $\chi^2$ divided by the number of data points \citep{skilling}. In its first series of iterations, the ZDI code produces magnetic models with associated synthetic profiles that progressively get closer to the target $\chi^2$ value. When the required reduced $\chi^2$ value is reached,
new iterations increase the entropy of the model (at fixed $\chi^2$), converging step by step towards the magnetic model that minimises the total information of the magnetic map. 

\subsection{Radial velocity}
The radial velocity of HN Peg was determined by fitting a Gaussian  directly to the Stokes I profile to determine the centroid of the profile. 
This method was applied to each epoch of HN Peg. Additionally the radial velocity in our ZDI code was varied in 0.1 kms$^{-1}$ steps.
The radial velocity which results in the minimum information content was chosen which was comparable to the
radial velocity obtained by  Gaussian fit. The radial velocity and the associated uncertainty for each observational epoch is shown in Table \ref{tablepercentage}.

\subsection{Inclination angle}
The inclination angle of 75$^\circ$ was inferred using the stellar parameters of HN Peg as shown in Table~\ref{table:1}, which was tested within its error range by using as an input to the ZDI code. The inclination angle was increased in 
5$^\circ$ steps and the inclination angle which resulted in minimum information content was used to generate the magnetic maps.

\subsection{Differential rotation}
 
The data used to reconstruct the magnetic field topology were collected over a span of several weeks, which might result into the introduction of latitudinal differential rotation during our timespan of observation. Differential rotation of HN Peg
was measured by determining the difference in equatorial and polar shear incorporating a simplified solar-like differential rotation law into the imaging process:

\begin{figure}
  \centering
 \includegraphics[width=.55\textwidth]{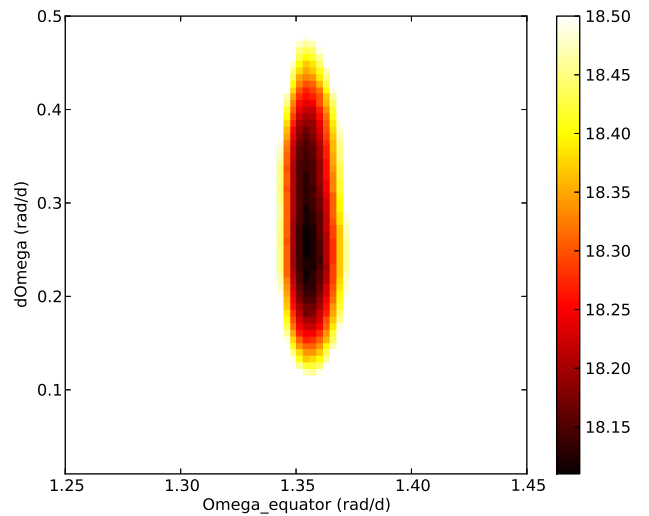}
 \caption{Best fit $\chi^2$ map obtained by varying the parameters for 2007 data. The $\Omega_\mathrm{eq}$ and d$\Omega$ values obtained from this map are  1.36 rad d$^{-1}$ and 0.22 rad d$^{-1}$ respectively.}
 \label{diffrotfit}
\end{figure}
 \begin{equation}
\Omega(l) = \Omega_\mathrm{eq}-d\Omega\sin^2l 
\label{diffrot}
\end{equation}
where $\Omega(l)$ is the rotation rate at latitude $\it l$, $\Omega_\mathrm{eq}$ is the equatorial rotation and $\it d\mathrm{\Omega}$ is the difference in rotation between the equator and the poles.

For a given set of 
$\Omega_\mathrm{eq}$ and $\it d\mathrm{\Omega}$ the large-scale magnetic field geometry was reconstructed, following the method of \citet{petit02}.
 Approximately 15 observations with good phase coverage \citep{morgenthaler12} is required to retrieve the parameters of the surface differential law. 
As good phase coverage is important to perform differential rotation calculations, the two epochs 2007.67 (14 observations) and 2013.68 (13 observations) were selected and individual differential rotation parameters 
were calculated. As shown in Fig~\ref{diffrotfit}, the $\Omega$ and d$\Omega$ values for 2007.67 epoch are 1.36$\pm$0.01 rad d$^{-1}$ and 0.22$\pm$0.03 rad d$^{-1}$ 
and for 2013.68 epoch are 1.27$\pm$0.01 rad d$^{-1}$ and 0.22$\pm$0.02 rad d$^{-1}$ respectively.
For the other epochs with less dense phase coverage the differential rotation values measured for 2007.67  are used for 2008.71, 2009.52 and values from 2013.68 are used for 2010.62 and 2011.67. 
The rotational period of HN Peg was also measured from the calculated differential rotational parameters.
\paragraph{}
The uncertainty in the differential rotation measurements were evaluated by obtaining $\Omega_\mathrm{eq}$ and d$\Omega$ values by varying the input stellar parameters, 
within the error bars of the individual parameters. The dispersion in the resulting 
values were considered as error bars.

\subsection{Magnetic topology}
\begin{figure*}
 \centering

   \includegraphics[width=.6\textwidth]{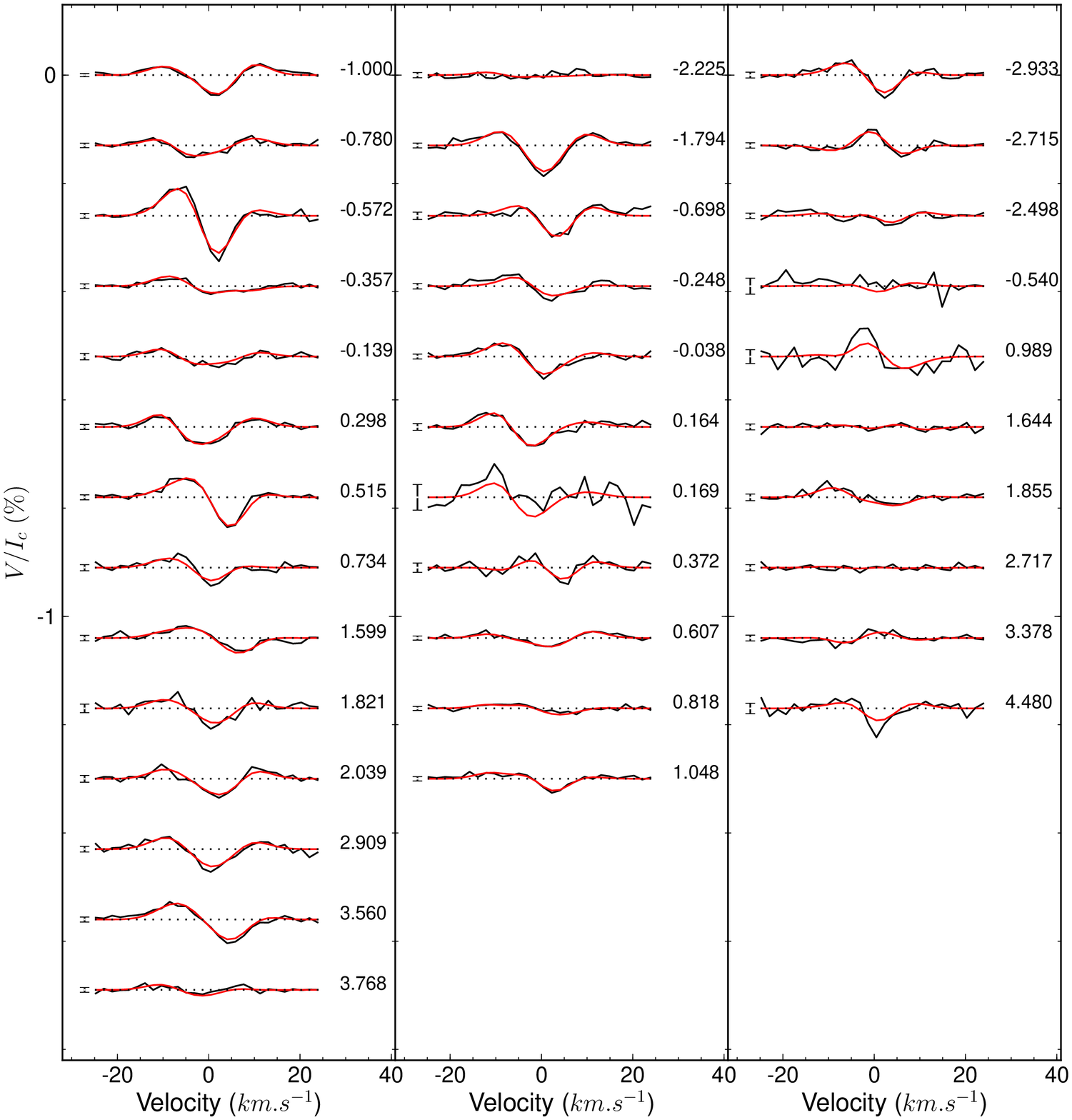}\\[5mm]
   \includegraphics[width=.6\textwidth]{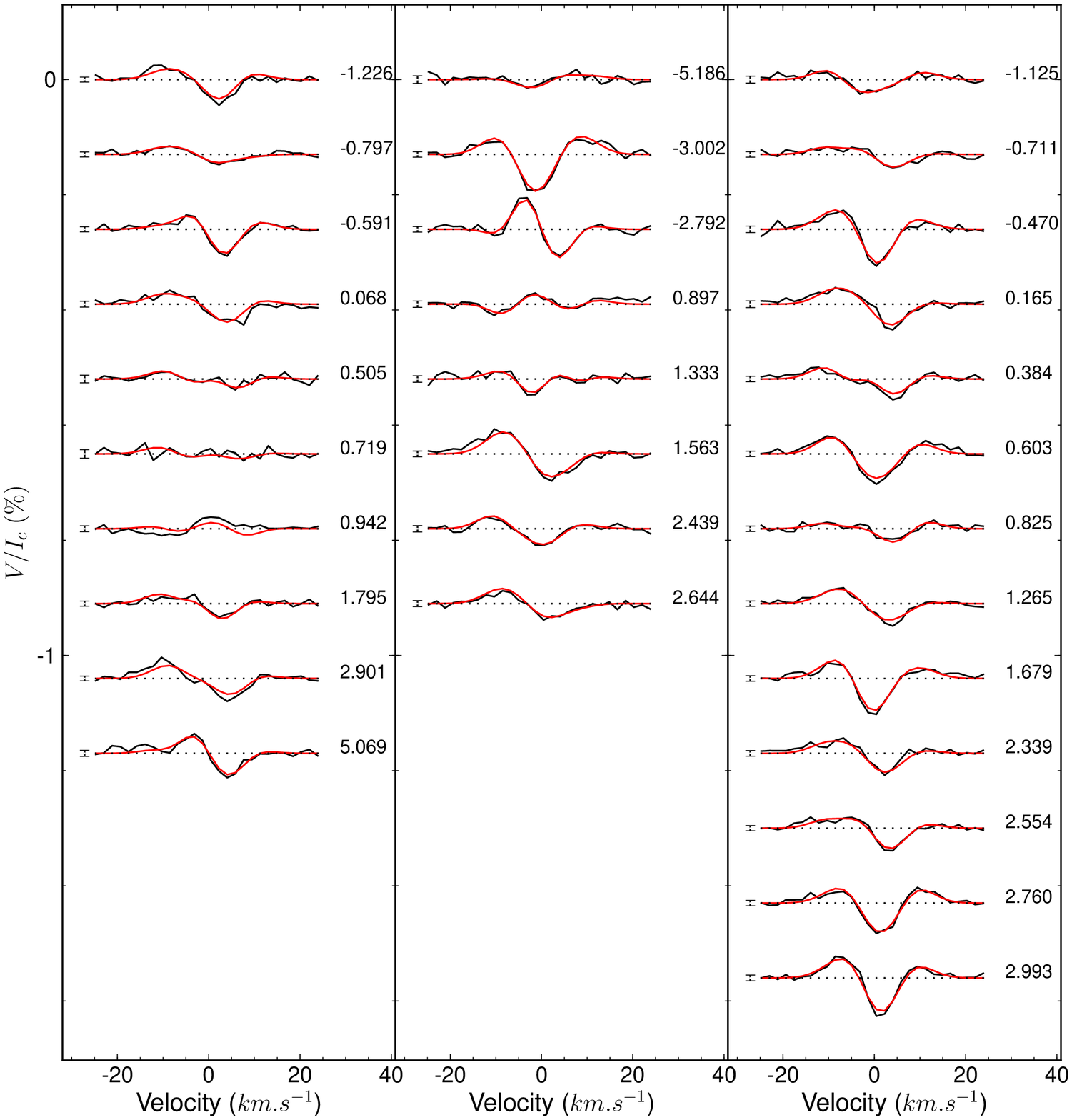}
   \caption{{\it Top row}: Time series of the LSD  Stokes V profiles from 2007.67 ({\it Top left}), 2008.71 ({\it Top centre}) and 2009.54 ({\it Top right}). {\it Bottom row}:
   Time series of the LSD Stokes V profiles for the epochs 2010.62 ({\it Bottom left}), 2011.67 ({\it Bottom centre}) and 2013.68 ({\it Bottom right}). The black line represents the observed Stokes V spectra and the red line 
   represents the fit to the spectra. Rotational cycle is shown to the right and 1$\sigma$ error bars for each observations is shown to the left for each plot.} 
     \label{timeseries}%
\end{figure*}

The large-scale magnetic field topology of HN Peg was reconstructed using ZDI, for the epochs 2007.67, 2008.71, 2009.54, 2010.62, 2011.67, and 2013.68. The stellar parameters used to reconstruct the magnetic field topology 
are a v$ \sin i$ of 10.6 kms$^{-1}$, an inclination angle of 75$^\circ$. The number of spherical harmonics $\ l$ used in our ZDI code is $\ l_{\mathrm {max}}$=8.  
\begin{figure*}
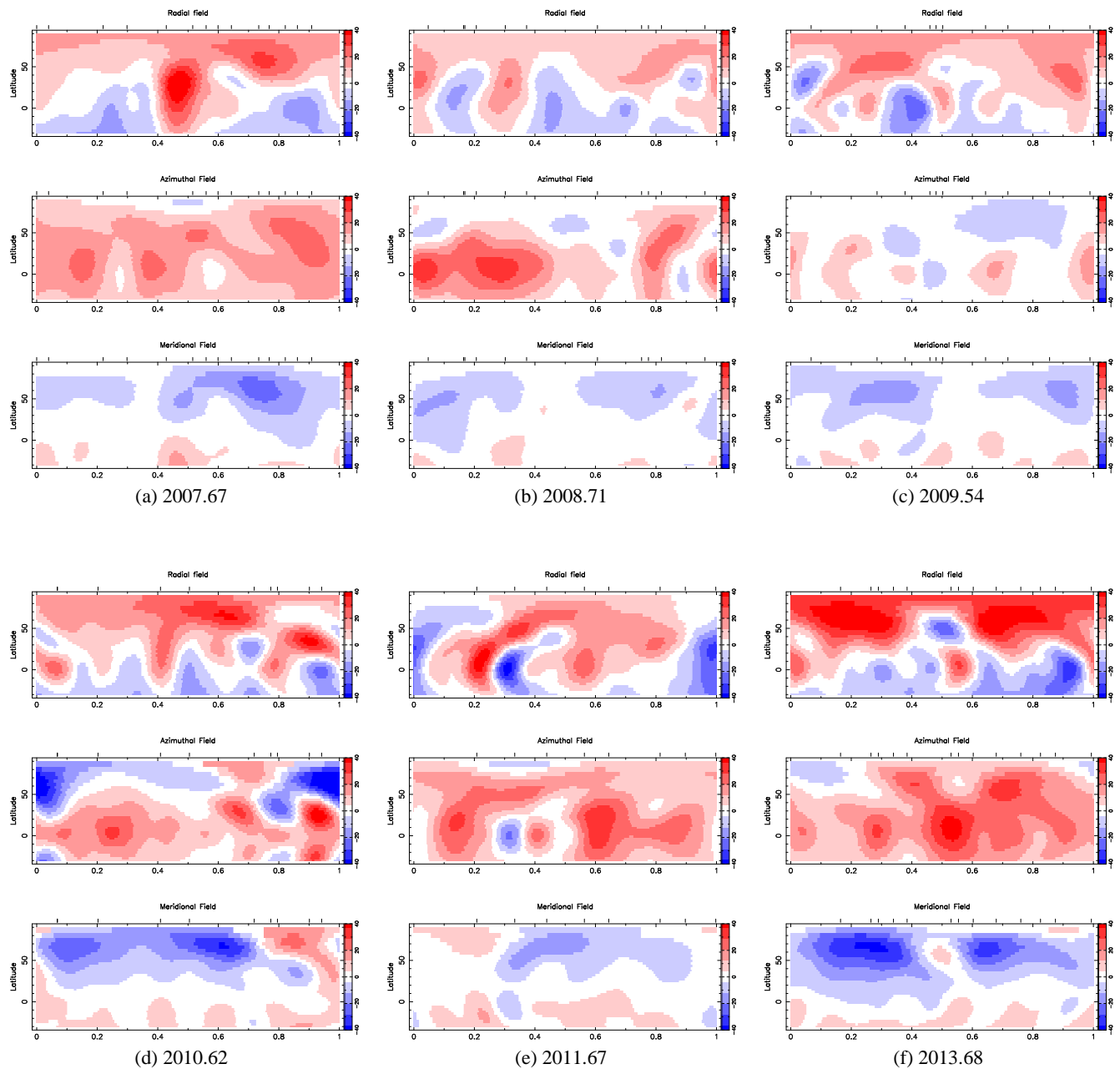

 \centering
\begin{tabular}{ccc}
\subfloat[2007.67] { \includegraphics[width=.3\textwidth]{ZDI07.eps}}&
\subfloat[2008.71]{\includegraphics[width=.3\textwidth]{ZDI08.eps}}&
\subfloat[2009.54]{\includegraphics[width = .3\textwidth]{ZDI09.eps}}\\[10mm]
\subfloat[2010.62] { \includegraphics[width=.3\textwidth]{ZDI10.eps}}&
\subfloat[2011.67]{\includegraphics[width=.3\textwidth]{ZDI11.eps}}&
\subfloat[2013.68]{\includegraphics[width = .3\textwidth]{ZDI13.eps}}\\
\end{tabular}  
   \caption{Surface magnetic field geometry of HN Peg for six epochs as reconstructed using Zeeman Doppler Imaging. $\it {Top}$ $\it row$: (a) 2007.67, (b) 2008.71, (c) 2009.54. $\it{ Bottom}$ $\it row$ : (d) 2010.62, (e) 2011.67, (f) 2013.68. 
   For each epoch, the magnetic
   field components are shown as projection onto one axis of the spherical coordinate frame, where from Top to Bottom: radial, azimuthal and meridional magnetic field components are shown. the field strength 
   is shown in Gauss, where red represents positive polarity and blue represents negative polarity.}
    \label{maps1}

    \end{figure*}

\subsubsection{Epoch 2007.67}
For the epoch 2007.67, the modelled Stokes V profile and the corresponding fit to the observed Stokes V profile is shown in Fig~\ref{timeseries} (Top left). The observed fit was achieved with a reduced  $\chi^2$ of 1.0. 
The number of degree of freedom is 152. In the radial field component of the magnetic field as shown in Fig~\ref{maps1}, a strong
positive field region is reconstructed at the equator along with a cap of positive polarity magnetic field encircling the pole. The azimuthal component is reconstructed
as a band of positive  magnetic field at equatorial latitudes as shown in Fig~\ref{maps1}. The percentage of the total magnetic energy distributed into its poloidal and toroidal configuration
for the epoch 2007.67 is shown in Fig~\ref{poltor}. The magnetic energy is 57$\%$ poloidal and 43$\%$ toroidal as shown in Table~\ref{tablepercentage}. The percentage of the poloidal field
reconstructed into its different components is shown in Fig~\ref{benergy}. The mean magnetic field strength of HN Peg is 18$\pm$0.5 G (Table~\ref{tablepercentage}). 

\subsubsection{Epoch 2008.71} 

The LSD Stokes V profile and the corresponding fit for the epoch 2008.71 is shown in Fig~\ref{timeseries} (Top middle). The observed fit was achieved with a
$\chi^2$ minimised to 1.0 and the number of degree of freedom is 68. The radial field geometry is dominated by a positive magnetic region over the poles as shown in Fig~\ref{maps1}, where the strong positive field at the equator in epoch 2007.67 is not
visible one year later in epoch 2008.71. The azimuthal field geometry is dominated by two regions 
of positive polarity at the equator. The meridional field geometry is also shown in Fig~\ref{maps1}. The percentage of the total magnetic energy distributed into the poloidal and toroidal components is shown in Fig~\ref{poltor}. 
The majority of the magnetic energy
is reconstructed as toroidal field component as shown in Table~\ref{tablepercentage}. The percentage of fraction of the poloidal magnetic field reconstructed into its different components is shown in Fig~\ref{benergy}.
The mean magnetic field strength decreases to 14$\pm$0.3 G(Table~\ref{tablepercentage}). 

\subsubsection{Epoch 2009.54}

The Stokes V profile in 2009.54 is shown in Fig~\ref{timeseries} (Top right), where some of the Stokes V profile have a lower S/N ratio. The Stokes V profiles are fitted to the reconstructed profile with
a $\chi ^2$ level of 1.0 and the number of degree of freedom is 40. The positive polarity magnetic region around the pole in the previous epochs is also present in 
2009.54 as shown in Fig~\ref{maps1}.
The band of positive polarity azimuthal field observed in the previous epochs is surprisingly absent in this epoch as shown in Fig~\ref{maps1}, with only 11$\%$ of the magnetic energy being toroidal.
The percentage of magnetic energy reconstructed as poloidal component is 89$\%$ as shown in Fig~\ref{poltor}.
The percentage of the fraction of the poloidal magnetic energy distributed into its different field configurations is shown in Fig~\ref{benergy}. 
The mean magnetic field strength is at its lowest at 11$\pm$0.2 G. (Table~\ref{tablepercentage}).

\subsubsection{Epoch 2010.62}

The Stokes V profile and magnetic maps of HN Peg for epoch 2010.62 are shown in Fig~\ref{timeseries} (Bottom left) and Fig~\ref{maps1} respectively. The best fit to the observed
Stokes V profiles were obtained with a $\chi ^2$ level of 1.4, which might be a result of some intrinsic behaviour that could not be accounted for. The number of degree of freedom is 40. The radial field component is still mostly positive around the poles as shown 
in Fig~\ref{maps1}. The azimuthal field is stronger than in epoch 2009.54, with the presence of  both positive and negative polarity regions. The meridional field component is also shown in Fig~\ref{maps1}. 
The percentage of the magnetic energy distribution into its poloidal and toroidal component is shown in Fig~\ref{poltor}. 65$\%$ of the magnetic energy is reconstructed in the poloidal component and 35$\%$ of the energy is
stored in the toroidal component as shown in Table~\ref{tablepercentage}. The percentage of the fraction of the poloidal component reconstructed into its different components is shown in
Fig~\ref{benergy}, where the fraction of the different components is minimal. 
The mean magnetic field strength of HN Peg has increased from 11$\pm$0.2 G in 2009.54 to 19$\pm$0.8 G (Table~\ref{tablepercentage}).

\subsubsection{Epoch 2011.67}
\begin{figure}
 \centering
 \includegraphics[width =.5\textwidth]{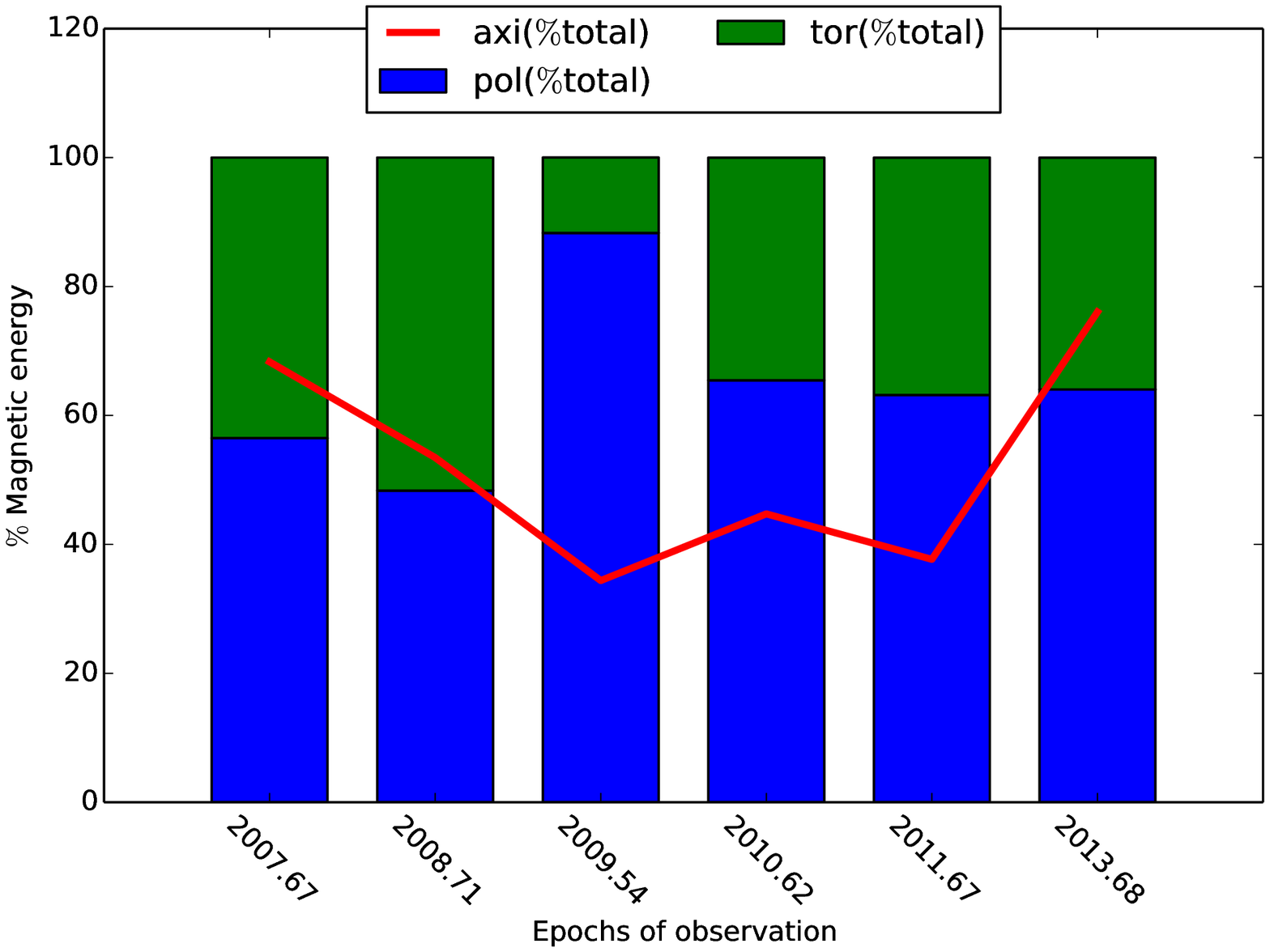}
 \caption{Magnetic energy distribution throughout the six epochs (2007.67, 2008.71, 2009.54, 2010.62, 2011.67, and 2013.68). 
 The fraction of magnetic field stored in poloidal component is shown in blue and toroidal component in green.The red line represents the fraction of the energy stored in the axisymmetric
 component. The error bars associated with each epoch are shown in Table~\ref{tablepercentage}.}
 \label{poltor}
\end{figure}

\begin{figure}
 \centering
 \includegraphics[width =.5\textwidth]{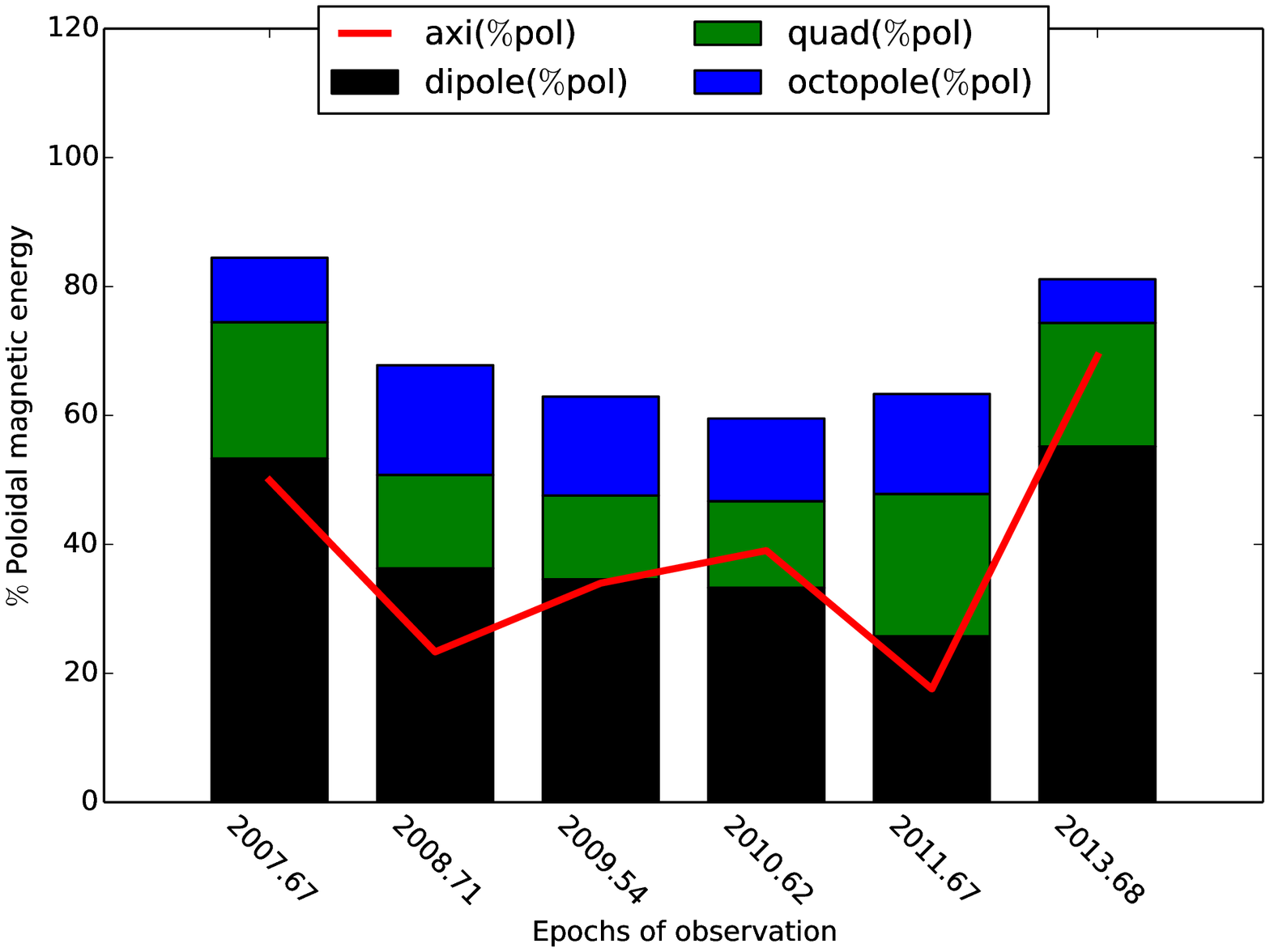}
 \caption{Poloidal magnetic field distributed into different configurations throughout the six epochs(2007.67, 2008.71, 2009.54, 2010.62, 2011.67, and 2013.68). 
 The fraction of the poloidal magnetic energy stored as dipole is shown in black, quadrupole in green and octopole in blue. The red line represents the fraction of the poloidal energy stored\
 in the axisymmetric component. The error bars associated with each epoch are shown in Table~\ref{tablepercentage}.}
 \label{benergy}
\end{figure}

For the epoch 2011.67, the reconstructed Stokes V profile and its fit to the observed Stokes V profile is shown in Fig~\ref{timeseries} (Bottom middle).  The observed fit was obtained with a
$\chi^2$ minimised to 1.1. The formal computation of the number of degree of freedom for this epoch
results in a negative value, which means for this particular case the problem is underdetermined.
The magnetic field geometry in the radial field is more complex than in the previous epochs, with the presence of both positive and negative magnetic regions as shown in 
Fig~\ref{maps1}. However, the phase coverage is not sufficient to reliably confirm the negative polarity regions.  
The azimuthal field is mostly positive with regions of positive polarity around the equator. The meridional field is  also shown in Fig~\ref{maps1}. The percentage of the magnetic energy distributed into the poloidal and toroidal 
components is shown in Fig~\ref{poltor}. The percentage of the fraction of the poloidal component reconstructed into its different field configurations is shown in Fig~\ref{benergy}.
61$\%$ of the magnetic energy is reconstructed into its poloidal component as shown in Table~\ref{tablepercentage}, where the mean magnetic field strength of HN Peg is 19$\pm$0.7 G.


\subsubsection{Epoch 2013.68}
The observed and reconstructed Stokes V profiles for the epoch 2013.68 is shown in Fig~\ref{timeseries} (Bottom right). The best fit to the observed Stokes V profiles was obtained with a $\chi^2$ of 1.0. The number of degree 
of freedom is 124. The magnetic field in the radial component is shown in Fig~\ref{maps1}, where the pole is dominated with a ring of positive polarity magnetic field. The azimuthal field component is dominated by a band of positive polarity magnetic field
at the equator as shown in Fig~\ref{maps1}. The percentage of the magnetic energy distributed into different field configurations is shown in Fig~\ref{poltor}.
The magnetic energy is mostly poloidal (62$\%$) as shown in Table~\ref{tablepercentage}. The percentage of the fraction of poloidal field reconstructed into its different configurations is shown in Fig~\ref{benergy}.
The mean magnetic field is at its highest at 24$\pm$0.7 G, where 77$\%$ of the total field is axisymmetric (Table~\ref{tablepercentage}).

\begin{table*}
\caption{Magnetic properties of HN Peg extracted from the ZDI maps. The columns represent fractional dates,number of observations, radial velocity (v$_r$),
mean magnetic field strength (B$_\mathrm{mean}$), fraction of magnetic energy reconstructed as the poloidal component, fraction of the large scale magnetic field energy reconstructed as
toroidal component, fraction of poloidal magnetic field stored as dipole, quadrupole and octopole, fraction of the total magnetic energy in the axisymmetric component of the magnetic field and finally
the differential rotation parameters $\Omega_\mathrm{eq}$ and d$\Omega$.}              
\label{tablepercentage}      
\centering                                      
\begin{tabular}{c c c c c c c c c c c c c c  }          
\hline                        
Dates&no of&v$_{r}$ &B$_{mean}$& pol&tor&dipole&quad&oct&axi&$\Omega_\mathrm{eq}$&d$\Omega$\\
(in frac.years)&obs&\textbf{(km s$^{-1}$)}&(G)&($\%$tot)&($\%$tot)&($\%$pol)&($\%$pol)&($\%$pol)&($\%$tot)&(rad d$^{-1}$)&(rad d$^{-1}$)\\
\hline                                   
2007.67&14&-16.60$\pm$0.17&18$\pm$0.5 &  57$\pm$4 & 43$\pm$24 & 54$\pm$4 & 22$\pm$4 & 9$\pm$1 & 69$\pm$15 & 1.36$\pm$0.01&0.22$\pm$0.03\\
2008.71&10&-16.67$\pm$0.04&14$\pm$0.3 & 49$\pm$12 & 51$\pm$16 & 35$\pm$3 & 15$\pm$2 & 17$\pm$1 & 54$\pm$21&--&--\\
2009.54&11&-16.63$\pm$0.20&11$\pm$0.2 &  89$\pm$23 & 11$\pm$11 & 42$\pm$4 & 11$\pm$3 & 14$\pm$1 & 45$\pm$21&--&--\\
2010.62&11&-16.65$\pm$0.02&19$\pm$0.8 &  65$\pm$13 & 35$\pm$10& 33$\pm$4 & 13$\pm$3 & 12$\pm$2 & 44$\pm$20&--&--\\
2011.67&8&-16.66$\pm$0.08&19$\pm$0.7 &  61$\pm$22 & 39$\pm$18 & 25$\pm$4 & 19$\pm$3 & 16$\pm$1& 38$\pm$21&--&--\\
2013.68&13&-16.66$\pm$0.03&24$\pm$0.7 & 62$\pm$4 & 38$\pm$24 &57$\pm$4 & 17$\pm$4 & 6$\pm$1& 77$\pm$10&1.27$\pm$0.01&0.22$\pm$0.02\\
    \hline                                             
\end{tabular}
\end{table*} 
\paragraph{}
The uncertainties associated with the magnetic maps for each epoch of observations were obtained by using different values for the input stellar parameters into our ZDI code \citep[see][]{petit02}, where the individual parameters were
varied within their error bars. The dispersion in the resulting values was considered as error bars.

\section{Discussion}
HN Peg was observed for seven epochs from 2007.67 to 2013.68, providing new insights into its magnetic field variations and the associated geometry.

\subsection{Long-term magnetic variability}
The longitudinal magnetic field (B$_l$) for each observation of HN Peg was derived using Stokes V profile and Stokes I profile integrated over the entire visible stellar surface. The longitudinal field varies as a function of the rotational
phase during each observational epoch, which indicates a non-axisymmetric magnetic geometry. 
Over the epochs of this analysis no significant long-term B$_l$ variations are apparent as shown in Fig~\ref{mean}, where the mean B$_{l}$ values exhibit variability over the observational timespan but the overall trend with 
dispersion is flat. HN Peg exhibits a strong longitudinal magnetic 
field strength when compared to the other solar type stars included in \citet{marsden}. B$_l$ ranges from -14 G up to 13 G (Table~\ref{values}) throughout the entire time span. 

\paragraph{}
The long term  chromospheric variability of
HN Peg was monitored using three different chromospheric lines: Ca II H$\&$K, H$\alpha$ and Ca II IRT. The three chromospheric tracers exhibit weak rotational dependence during each observational epoch. 
Periodic analysis of HN Peg carried out by the Mount Wilson survey categorised HN Peg as a variable star with a period of 6.2 $\pm$ 0.2 years. The chromospheric activity of HN Peg was also measured as part of 
the Bcool snapshot survey \citep{marsden}, where the measured S-index are compatible with our S-index measurements.

\paragraph{} 
Correlations were observed for the three chromospheric tracers  for individual epochs with visible scatter, which might be due to the effect of different temperature and pressure conditions during emissions in the line cores 
of Ca II H$\&$K, H$\alpha$, and Ca II IRT. Correlations between S-index and H$\alpha$-index were also observed for $\xi$ Bootis A by \citet{morgenthaler12}. The observed scatter in the correlation between S-index and H$\alpha$-index for each epoch might be also be explained by the contribution of plage variation in Ca II H$\&$K and 
filament variation in H$\alpha$ flux \citep{meunier}, where increase in filament contribution might result in decrease in correlation between the two chromospheric tracers. Apart from the contribution of 
filament and plage,
\citet{meunier} also concluded that stellar inclination angle, phase coverage might also effect the correlation between these two tracers.
\paragraph{}
In their long-term evolution, the chromospheric tracers exhibit a similar trend, with visible correlation 
between the S-index and H$\alpha$-index. In the long-term mean S-index and H$\alpha$-index exhibit correlation in cool stars, which might be due to the effect of stellar colour \citep{cincunegui}.   
The long-term Ca II H$\&$K and H$\alpha$ correlation is clearly observed in the Sun \citep{livingston}, where the two activity proxies follow the solar magnetic
cycle. This long-term correlation between these two tracers have also been observed in other cool stars with high activity index \citep{gomes}. 
\paragraph{}
For each observational epoch no visible correlation is observed between the variability of the longitudinal magnetic field and the 
chromospheric tracers.  
Correlations between the direct field measurements and the magnetic activity proxies were also  not observed for the solar analogue $\xi$ Bootis A \citep{morgenthaler12}.  
This lack of correlation can be explained by the contribution of small scale magnetic features in chromospheric activity measurements which are lost in the polarised Stokes V magnetic field calculations due to 
magnetic flux cancellations.

\subsection{Large scale magnetic topology}

The large-scale magnetic topology of HN Peg was reconstructed for six observational epochs (2007.67, 2008.71, 2009.54, 2010.62, 2011.67, and 2013.68), where the mean magnetic field strength (B$_\mathrm{mean}$) changes
with the geometry of the field from epoch-to-epoch. The mean magnetic field strength of HN Peg is of a few G, which is considerably smaller when compared to the mean magnetic field of
other solar analogues HD 189733(1.34$\pm$0.13M$_{\sun}$, T$_\mathrm{eff}$=6014 K) \citep{hd189733} and $\xi$ Bootis A (0.86$\pm$0.07 M$_{\sun}$, T$_\mathrm{eff}$=5551$\pm$20) \citep{morgenthaler12}. The mean magnetic field strength of HN Peg is higher than HD 190771, which 
has a mass of 0.96$\pm$0.13 M$_\sun$ and T$_\mathrm{eff}$ of 5834$\pm$50 \citep{petit09,morgenthaler11}.
When compared to solar-type stars of similar spectral type HN Peg exhibits higher mean field strength than the mean field strength of $\tau$ Boo(F7) with a mass of 1.33$\pm$0.11 M$_\sun$ \citep{fares}, HD 179949(F8) \citep{hd179949}, 
where $\tau$ Boo and 
HD 179949 are both planet hosting stars. 


\paragraph{}
The radial field component of HN Peg exhibits a variable field geometry, where the field strength varies from epoch-to-epoch as shown in Fig~\ref{maps1}.  
A positive polarity region at the pole is observed in epoch 2007.67, where a strong positive polarity magnetic region is also observed near the equator. The positive region at the pole is present throughout 
our observational epoch, with out exhibiting any polarity switch. The magnetic field energy is stored into its poloidal and toroidal components.
The poloidal field of HN Peg is not a simple dipole.
The fraction of energy stored into the different components of the poloidal field exhibits variations from epoch-to-epoch as shown in Fig~\ref{benergy}.

\paragraph{}
The azimuthal field component of HN Peg exhibits a more variable geometry compared to the radial field geometry. The azimuthal field component exhibits the presence of a significant positive polarity
magnetic regions, which undergoes variations from epoch-to-epoch as shown in Fig~\ref{maps1}. A strong positive polarity  band of magnetic field encircling the star is observed in epoch 2007.67. In 2008.71 two strong
positive polarity magnetic regions were observed near the equator. The azimuthal component becomes negligible in the epoch 2009.54. The toroidal field percentage is minimum in epoch 2009.54. The azimuthal field reappears 
in 2010.62, where opposite polarity magnetic field regions are observed. In 2011.67 stronger positive polarity regions are observed, which finally appears as a toroidal ring in epoch 2013.68.  

\paragraph{}
Prominent toroidal features have been observed in a wide range of stars belonging to different spectral class, such as HD 190771 
\citep{petit09}, $\xi$ Bootis A \citep{morgenthaler12}, $\tau$ Boo \citep{fares} and HD 189733 \citep{hd189733}. The sudden disappearance of the toroidal field was also observed in $\xi$ Boo \citep{morgenthaler12}. 
The toroidal component is prominent in solar-type stars with rotation periods as short as a few days \citep{pascal08} and stars  with longer rotational periods show more prominent poloidal component,
which is clearly observed in the Sun. Toroidal band was also not observed in the F8 dwarf HD 179949 \citep{hd179949}, where only two epochs of observations were available. The presence of significant global-scale toroidal field has 
also been observed in numerical simulations of rapidly rotating suns \citep{brown10}, where the surface field topology becomes predominantly toroidal for stars with rotation periods of a few days.

\paragraph{}
No polarity switches have been observed
for HN Peg, although it showed significant evolution of its magnetic field geometry over the span of six observational epochs.  Magnetic cycles were also not observed for $\xi$ Bootis A and HD 189733, which are 
slower rotators when compared to HN Peg.
Polarity switches were observed in HD 190771, $\tau$ Boo and HD 78366 over their observational time span. Magnetic cycles shorter than the magnetic cycle of the Sun were observed for $\tau$Boo
and HD 78366. 

\paragraph{} 
The variability of the mean magnetic field of HN Peg follows a similar trend as that of the toroidal field component (Table~\ref{tablepercentage}).
The mean magnetic 
field (B$_\mathrm{mean}$) of HN Peg show a gradual decrease in its field strength from 2007.67 to 2009.54, with minimum B$_\mathrm{mean}$ in 2009.54. The mean field starts increasing 
from 2009.54 till it reaches a maximum strength in 2013.68. This indicates strong dependence of the mean field strength on the toroidal component of the magnetic field.  

\subsection{Differential rotation}

HN Peg has a low $\it{v\sin i}$ of 10.6 kms$^{-1}$, which makes it unsuitable for differential rotation 
calculations using other conventional techniques such as line profile studies using Fourier Transform method  \citep{reiners}. Photometric observations of HN Peg were used to measure its differential 
rotation by \citet{antisolar}, where the evolution of the rotation of the star along the star spot cycle was measured with inconclusive results. 
\paragraph{}

The differential rotation of HN Peg was calculated using Stokes V and I profiles in the ZDI technique. 
Two epochs with the best phase coverage were used (2007.67 and 2013.68) in our differential rotation calculations.
The $\Omega$ and d$\Omega$ values for 2007.67 epoch were 1.36$\pm$0.01 rad d$^{-1}$ and 0.22$\pm$0.03 rad d$^{-1}$ and for 2013.68 epoch were 1.27$\pm$0.01 rad d$^{-1}$ and 0.22$\pm$0.02 rad d$^{-1}$ respectively.
\paragraph{}

HN Peg exhibits weak differential rotation compared to other dwarfs of similar spectral types such as HD 171488 (G0) (StokesI/StokesV: $\Omega_\mathrm{eq}$=4.93$\pm$0.05/4.85$\pm$0.05 rad d$^{-1}$, d$\Omega$=0.52$\pm$
0.04/0.47$\pm$0.04 rad d$^{-1}$) \citep{sjeffers} and $\tau$Boo (F7) (2008 June: $\Omega_\mathrm{eq}$=2.05$\pm$0.04 rad d$^{-1}$ and d$\Omega$=0.42$\pm$0.10 rad d$^{-1}$ rad d$^{-1}$, 
2008 July: $\Omega_\mathrm{eq}$=2.12$\pm$0.12 rad d$^{-1}$ and d$\Omega$=0.50$\pm$0.15 rad d$^{-1}$) \citep{fares}.  
HD 171488 is the closest to HN Peg in terms of spectral type, stellar radius and age.
The d$\Omega$ values of HN Peg is higher than the other young early G dwarfs such as LQ Lup($\Omega_\mathrm{eq}$=20.28$\pm$0.01 rad d$^{-1}$, d$\Omega$=0.12$\pm$0.02 rad d$^{-1}$) \citep{donati2000} and 
R58 (2000 January:$\Omega_\mathrm{eq}$=11.14$\pm$0.01 and d$\Omega$=0.03$\pm$0.02, 2003 March: $\Omega_\mathrm{eq}$=11.19$\pm$0.01 and d$\Omega$=0.14$\pm$0.01) \citep{marsden2004}.
When compared to HD 179949($\Omega_\mathrm{eq}$=0.82$\pm$0.01 rad d$^{-1}$, d$\Omega$=0.22$\pm$0.06 rad d$^{-1}$)\citep{hd179949}, HN Peg exhibits comparable d$\Omega$ values. 
Although, when compared to other fast rotators such as $\xi$ Boo (P$_\mathrm{rot}$=6.43 days) \citep{morgenthaler12}, HN Peg exhibits weaker differential rotation. 
No direct correlation between differential rotation  of HN Peg and  solar analogues of similar stellar parameters such as age, spectral type, P$_\mathrm{rot}$ have been observed so far. 

\section{Summary}
In this paper we presented the large-scale magnetic topology of the young solar analogue HN Peg.  
HN Peg is a variable young dwarf with a complex magnetic geometry, where the radial field exhibits stable positive polarity magnetic field region through out our observational epochs. In contrast, the azimuthal field exhibits a
highly variable geometry where a band of positive polarity toroidal field is observed in the first epoch of observation followed by a negligible toroidal field two years later in epoch 2009.54. The toroidal band emerges again one
year later in epoch 2010.62 which is stable in the later epochs 2011.67 and 2013.68. The long-term longitudinal magnetic field variations were also calculated where in the long-term the longitudinal field exhibits a flat trend.
The chromospheric activity was also measured, where the chromospheric activity indicators exhibit a long-term correlation.

\begin{acknowledgements}
 This work was carried out as part of Project A16 funded by  the Deutsche Forschungdgemeinschaft (DFG) under SFB 963. Part of this work was also supported by the COST Action MP1104 
 "Polarisation as a tool to study the Solar System and beyond".
\end{acknowledgements}

 \begin{longtab}
\begin{longtable}{cccccc}
\caption{\label{kstars} Journal of observations for seven epochs(2007-2013). Column 1 represents the year and date of observations, column 2 is the Heliocentric Julian date, 
column 3 is the exposure time, column 4 is the signal-to-noise ratio of each Stokes V LSD profile and column 5 represents the error bars in Stokes V LSD profile.}\\
\hline\hline
Date & Julian date & Exposure time &S/N &$\sigma_\mathrm{LSD}$\\
&(2454000+)& (s)& & (10$^{-5} \mathrm {I_{c}} $) \\
\hline
\endfirsthead
\caption{continued.}\\
\hline\hline

Date & Julian date &Exposure time & S/N&$\sigma_\mathrm{LSD}$\\
&(2454000+)&(s)&&(10$^{-5} \mathrm{I_{c}}$) \\
\hline
\endhead
\hline
\endfoot

2007 July 27&309.595010 & 2400& 33888&2.9509 \\
2007 July 28&310.602470 & 1200& 24226&4.1279 \\
2007 July 29&311.560380 & 1200&22030&4.5393\\
2007 July 30&312.547230&1200&23060&4.3366\\
2007 July 31 &313.548380&1200&17034&5.8706\\
2007 August 2& 315.554720&1200&21801&4.5871\\        
2007 August 3&316.551130&1200&21899&4.5664\\
2007 August 4&317.552700&1200& 17688&5.6536\\
2007 August 8&321.525450&1200&20126&4.9688\\
2007 August 9&322.545370&1200&12680&7.8868\\
2007 August 10&323.545440&1200&15856&6.3068\\
2007 August 13&326.520920&1200&2225&0.0004\\
2007 August 14&327.534930&1200&19289&5.1843\\
2007 August 17&330.524710&1200&21711&4.6061\\
2007 August 18&331.481500&1200&22426&4.4592\\
2007 August 26&339.432390&1200&3523&0.0003\\
\hline
2008 August 10&689.532480&1200&20933&4.7772\\
2008 August 12&691.512640&1200&22440&4.4563\\
2008 August 17&696.541340&1200&14770&6.7706\\
2008 August 19&698.608480&1200&19645&5.0903\\
2008 August 20&699.569970&1200&22309&4.4826\\
2008 August 21&700.499080&1200&16068&6.2235\\
2008 August 21&700.51924&1200&4218&0.0002\\
2008 August 22&701.454560&1200&11621&8.6048\\
2008 August 23&702.529090&1200&26050&3.8388\\
2008 August 24&703.500480&1200&26143&3.8251\\
2008 August 25&704.553140&1200&25206&3.9673\\
\hline
2009 June 1 &984.634140&1200&18628&5.3684\\
2009 June 2&985.633550&1200&20178&4.9560\\
2009 June 3&986.630640&1200&20354&4.9131\\
2009 June 12&995.615700&1200&6753&0.0001\\
2009 June 18&1001.610390&1200&2183&0.0005\\
2009 June 19&1002.632590&1200&7840&0.0001\\
2009 June 22&1005.639920&1200&18545&5.3923\\
2009 June 23&1006.610200&1200&16505&6.0586\\
2009 June 27&1010.568540&1200&17878&5.5936\\
2009 June 30&1013.598260&1200&18632&5.3672\\
2009 July 5&1018.658940&1200&10351&9.6608\\
\hline
2010 June 21&1369.590720&1200&18367&5.4444\\
2010 July 4&1382.615660&1200&23145&4.3206\\
2010 July 6&1384.584890&1200&23052&4.3381\\
2010 July 7&1385.533040&1200&22897&4.3674\\
2010 July 10&1388.555690&1200&19426&5.1477\\
2010 July 12&1390.561220&1200&15164&6.5946\\
2010 July 13&1391.543670&1200&13473&7.4222\\
2010 July 14&1392.566570&1200&19724&5.0700\\
2010 July 18&1396.485880&1200&22411&4.4621\\
2010 July 23&1401.561040&1200&23969&4.1721\\
2010 August 2&1411.510430&1200&19422&5.1488\\
2010 August 7&1416.561860&1200&21276&4.7002\\
2010 August 20&1429.547050&1200&15018&6.6587\\
\hline
2011 July 11&1754.590060&1200&15272&6.5481\\
2011 July 21&1764.613990&1200&21317&4.6911\\
2011 July 22&1765.579850&1200&21091&4.7414\\
2011 August 8&1782.512460&1200&17905&5.5850\\
2011 August 10&1784.515440&1200&17964&5.5668\\
2011 August 11&1785.571320&1200&19745&5.0646\\
2011 August 15&1789.590600&1200&18044&5.5420\\
2011 August 16&1790.530870&1200&19198&5.2088\\
2011 August 17&1791.472450&1200&--&--\\
2011 August 18&1792.535280&1200&--&--\\
\hline
2012 June 21&2100.610620&1200&--&--\\
2012 June 22&2101.635920&1200&--&--\\
2012 June 23&2102.612400&1200&--&--\\
2012 June 24&2103.529510&1200&--&--\\
2012 July 9&2118.614210&1200&--&--\\
2012 July 15&2124.664630&1200&--&--\\
2012 July 16&2125.639690&1200&--&--\\
2012 July 17&2126.631180&1200&--&--\\
2012 July 18&2127.593570&1200&--&--\\
2012 July 19&2128.578110&1200&--&--\\
2012 July 22&2131.572610&1200&--&--\\
2012 July 23&2132.561810&1200&--&--\\
2012 July 24&2133.592020&1200&--&--\\
2012 August 6&2146.566500&1200&--&--\\
2012 August 7&2147.537680&1200&--&--\\
\hline
2013 July 8&2482.556520&1200&21586&4.6327\\
2013 July 11&2485.512210&1200&14989&6.6715\\
2013 August 2&2507.629120&1200&16687&5.9928\\
2013 August 4&2509.533370&1200&21228&4.7107\\
2013 August 5&2510.636070&1200&17799&5.6183\\
2013 August 8&2513.550070&1200&17224&5.8058\\
2013 August 9&2514.559260&1200&20757&4.8178\\
2013 August 10&2515.562810&1200&21317&4.6912\\
2013 August 11&2516.582590&1200&21314&4.6917\\
2013 August 13&2518.601950&1200&21551&4.6401\\
2013 August 15&2520.501260&1200&20952&4.7728\\
2013 August 18&2523.532240&1200&20357&4.9122\\
2013 August 19&2524.518410&1200&22046&4.5360\\
2013 August 20&2525.463280&1200&21075&4.7450\\
2013 August 21&2526.533220&1200&19830&5.0429\\
\hline
\footnote{The spectropolarimetric observations are not included for 2012.61 and parts of 2011.67 epoch due to instrumental defects at NARVAL. See Sect.~3 for details.} 

 \end{longtable}
\end{longtab}

\begin{longtab}
\begin{longtable}{ccccccc}
\caption{\label{values} The chromospheric activity measurements and magnetic field measurements of HN Peg for seven epochs (2007.67, 2008.71, 2009.54, 2010.62, 2011.67, 2012.61 and 2013.68). 
From left to right it represents: Julian date, rotational phase, S-index, H${\alpha}$-index, Ca$\mathrm{IRT}$-index, longitudinal magnetic field (B$_{l}$) and magnetic field of the Null profile (N$_{l}$).}\\
\hline\hline

Julian date&rot.phase&S-index&H$\alpha$-index&Ca$\mathrm{IRT}$-index&B$_{l}$&N$_{l}$\\
(2454000+)&&&&&(G)&(G)\\
\hline
\endfirsthead
\caption{continued.}\\
\hline\hline

Julian date& rot.phase&S-index& H$\alpha$-index& Ca$\mathrm{IRT}$-index& B$_{l}$& N$_{l}$\\
(2454000+)&&&&&(G)&(G)\\
\hline
\endhead
\hline
\endfoot
309.595010&0.000000&0.3092$\pm$0.0032&0.3234$\pm$0.0008&0.9297$\pm$0.0044&1.8$\pm$0.5&-0.4$\pm$0.5\\
310.602470&0.219490&0.3158$\pm$0.0044&0.3276$\pm$0.0011&0.9384$\pm$0.0062&-1.6$\pm$0.7&0.8$\pm$0.7\\
311.560380&0.428185&0.3088$\pm$0.0048&0.3263$\pm$0.0012&0.9301$\pm$0.0067&12.7$\pm$0.8&0.0$\pm$0.8\\
312.547230&0.643185&0.3106$\pm$0.0045&0.3262$\pm$0.0012&0.9312$\pm$0.0065&5.1$\pm$0.7&-1.1$\pm$0.7\\
313.548380&0.861301&0.3099$\pm$0.0065&0.3248$\pm$0.0015&0.9286$\pm$0.0083&3.6$\pm$1.5&0.7$\pm$1.5\\
315.554720&0.298412&0.3162$\pm$0.0052&0.3273$\pm$0.0012&0.9341$\pm$0.0068&1.6$\pm$1.9&0.7$\pm$1.9\\
316.551130&0.515495&0.3104$\pm$0.0051&0.3255$\pm$0.0012&0.9287$\pm$0.0067&11.1$\pm$0.8&0.0$\pm$0.8\\
317.552700&0.733702&0.3086$\pm$.0063&0.3248$\pm$0.0015&0.9237$\pm$0.0084&7.4$\pm$1.3&-0.5$\pm$1.3\\
321.525450&0.599224&0.3082$\pm$0.0056&0.3254$\pm$0.0013&0.9287$\pm$0.0072&7.9$\pm$0.8&0.7$\pm$0.8\\
322.545370&0.821429&0.3045$\pm$0.0086&0.3263$\pm$0.0020&0.9241$\pm$0.0112&4.7$\pm$1.3&0.9$\pm$1.3\\
323.545440&0.039309&0.3097$\pm$0.0072&0.3282$\pm$0.0016&0.9284$\pm$0.0088&2.6$\pm$1.1&-0.4$\pm$1.1\\
327.534930&0.908479&0.4618$\pm$0.1629&0.3402$\pm$0.0118&0.9293$\pm$0.0633&3.8$\pm$0.9&1.2$\pm$0.9\\
330.524710&0.559847&0.3102$\pm$0.0062&0.3252$\pm$0.0014&0.9293$\pm$0.0075&12.2$\pm$0.8&0.5$\pm$0.8\\
331.481500&0.768298&0.3112$\pm$0.0051&0.3252$\pm$0.0012&0.9230$\pm$0.0067&1.8$\pm$1.2&0.7$\pm$1.2\\
\hline
689.532480&0.775048&0.3076$\pm$0.0048&0.3259$\pm$0.0012&0.9220$\pm$0.0065&-2.4$\pm$0.8&-0.7$\pm$0.8\\
691.512640&0.206455&0.3508$\pm$0.0689&0.3271$\pm$0.0073&0.7714$\pm$0.0437&-2.6$\pm$0.8&1.0$\pm$0.8\\
696.541340&0.302033&0.3080$\pm$0.0053&0.3216$\pm$0.0012&0.9110$\pm$0.0068&2.3$\pm$1.1&0.6$\pm$1.1\\
698.608480&0.752390&0.3285$\pm$0.0051&0.3269$\pm$0.0012&0.9254$\pm$0.0065&6.0$\pm$0.9&-1.0$\pm$0.9\\
699.569970&0.961865&0.3299$\pm$0.0085&0.3265$\pm$0.0017&0.9228$\pm$0.0092&7.0$\pm$0.8&-0.8$\pm$0.8\\
700.499080&0.164285&0.3145$\pm$0.0062&0.3244$\pm$0.0013&0.9129$\pm$0.0067&4.1$\pm$4.7&1.8$\pm$1.2\\
700.519240&0.168678&0.3075$\pm$0.0051&0.3233$\pm$0.0012&0.9069$\pm$0.0064&8.4$\pm$1.2&-5.3$\pm$4.7\\
701.454560&0.372451&0.3122$\pm$0.0692&0.3225$\pm$0.0015&0.9249$\pm$0.0084&-3.7$\pm$3.6&4.5$\pm$3.6\\
702.529090&0.606553&0.3041$\pm$0.0163&0.3230$\pm$0.0030&0.9304$\pm$0.0165&-1.3$\pm$1.1&-0.2$\pm$1.1\\
703.500480&0.818185&0.3233$\pm$0.0098&0.3275$\pm$0.0021&0.9201$\pm$0.0117&3.1$\pm$0.6&-0.4$\pm$0.6\\
704.553140&0.047523&0.3213$\pm$0.0043&0.3243$\pm$0.0010&0.9109$\pm$0.0056&3.7$\pm$0.7&0.9$\pm$0.7\\
\hline
984.634100&0.067340&0.3144$\pm$0.0042&0.3231$\pm$0.0010&0.9077$\pm$0.0056&5.3$\pm$0.9&0.1$\pm$0.9\\
985.633550&0.285085&0.3075$\pm$0.0043&0.3230$\pm$0.0011&0.9016$\pm$0.0058&1.7$\pm$0.8&0.3$\pm$0.8\\
986.630640&0.502316&0.3228$\pm$0.0063&0.3306$\pm$0.0013&0.9189$\pm$0.0070&5.1$\pm$1.6&2.6$\pm$1.6\\
995.615700&0.459845&0.3155$\pm$0.0056&0.3290$\pm$0.0012&0.9158$\pm$0.0065&8.5$\pm$4.4&-10.9$\pm$4.5\\
1001.610390&0.765878&0.3303$\pm$0.0059&0.3255$\pm$0.0012&0.9230$\pm$0.0066&-14.0$\pm$7.8&-17.7$\pm$7.8\\
1002.632590&0.988580&0.3036$\pm$0.0206&0.3245$\pm$0.0031&0.9098$\pm$0.0158&7.9$\pm$2.5&-0.7$\pm$2.5\\
1005.639920&0.643771&0.3107$\pm$0.0202&0.3250$\pm$0.0033&0.9318$\pm$0.0175&3.5$\pm$2.0&2.6$\pm$2.0\\
1006.610200&0.855161&0.3209$\pm$0.0180&0.3259$\pm$0.0029&0.9319$\pm$0.0154&5.9$\pm$1.0&-0.1$\pm$1.0\\
1010.568540&0.717545&0.3326$\pm$0.0080&0.3260$\pm$0.0013&0.9327$\pm$0.0072&-2.0$\pm$2.1&-1.5$\pm$2.1\\
1013.598260&0.377614&0.3338$\pm$0.0084&0.3271$\pm$0.0015&0.9259$\pm$0.0080&-4.5$\pm$1.8&-0.0$\pm$1.8\\
1018.658940&0.480159&0.3381$\pm$0.0072&0.3252$\pm$0.0014&0.9332$\pm$0.0072&3.1$\pm$1.7&-3.5$\pm$1.7\\
\hline
1369.590720&0.935885&0.3212$\pm$0.0074&0.3230$\pm$0.0013&0.9292$\pm$0.0068&13.9$\pm$0.9&0.3$\pm$0.9\\
1382.615660&0.773562&0.3193$\pm$0.0081&0.3251$\pm$0.0019&0.9147$\pm$0.0096&7.3$\pm$0.7&-0.7$\pm$0.7\\
1384.584890&0.202588&0.3577$\pm$0.0073&0.3291$\pm$0.0013&0.9423$\pm$0.0069&4.7$\pm$0.7&-0.7$\pm$0.7\\
1385.533000&0.409148&0.3323$\pm$0.0051&0.3255$\pm$0.0011&0.9265$\pm$0.0061&6.1$\pm$0.7&0.5$\pm$0.7\\
1388.555690&0.067686&0.3493$\pm$0.0057&0.3284$\pm$0.0011&0.9325$\pm$0.0062&9.5$\pm$1.0&-0.8$\pm$1.0\\
1390.561220&0.504621&0.3474$\pm$0.0056&0.3264$\pm$0.0011&0.9293$\pm$0.0062&3.2$\pm$1.1&0.9$\pm$1.1\\
1391.543670&0.718662&0.3446$\pm$0.0067&0.3256$\pm$0.0013&0.9293$\pm$0.0068&4.7$\pm$2.8&-0.8$\pm$2.8\\
1392.566500&0.941501&0.3370$\pm$0.0090&0.3262$\pm$0.0016&0.9263$\pm$0.0080&-4.6$\pm$1.0&2.1$\pm$1.0\\
1396.485880&0.795397&0.3290$\pm$0.0100&0.3244$\pm$0.0018&0.9133$\pm$0.0096&4.4$\pm$0.9&1.0$\pm$0.9\\
1401.561040&0.901096&0.3358$\pm$0.0061&0.3248$\pm$0.0013&0.9214$\pm$0.0068&10.0$\pm$0.7&1.1$\pm$0.7\\
1411.510430&0.068719&0.3314$\pm$0.0062&0.3264$\pm$0.0011&0.9156$\pm$0.0062&8.4$\pm$1.0&-0.3$\pm$1.0\\
1416.561860&0.169248&0.3302$\pm$0.0052&0.3262$\pm$0.0011&0.9155$\pm$0.0059&5.5$\pm$0.8&0.8$\pm$0.8\\
1429.547050&0.998266&0.3474$\pm$0.0072&0.3289$\pm$0.0013&0.9177$\pm$0.0069&8.1$\pm$1.5&0.1$\pm$1.5\\
\hline
1754.590060&0.813736&0.3520$\pm$0.0068&0.3287$\pm$0.0012&0.9170$\pm$0.0064&-2.3$\pm$1.1&-0.4$\pm$1.1\\
1764.613990&0.997599&0.3397$\pm$0.0135&0.3264$\pm$0.0017&0.9093$\pm$0.0095&-2.4$\pm$0.9&-0.1$\pm$0.9\\
1765.579850&0.208026&0.3392$\pm$0.0075&0.3281$\pm$0.0016&0.9445$\pm$0.0088&6.1$\pm$0.8&0.5$\pm$0.8\\
1782.512460&0.897048&0.3445$\pm$0.0040&0.3286$\pm$0.0012&0.9421$\pm$0.0068&-6.4$\pm$1.9&0.1$\pm$1.9\\
1784.515440&0.333427&0.3398$\pm$0.0048&0.3285$\pm$0.0012&0.9415$\pm$0.0068&4.3$\pm$2.1&1.4$\pm$2.1\\
1785.571320&0.563466&0.3475$\pm$0.0056&0.3289$\pm$0.0014&0.9458$\pm$0.0079&12.8$\pm$0.8&-1.0$\pm$0.8\\
1789.590600&0.439126&0.3292$\pm$0.0072&0.3278$\pm$0.0014&0.9324$\pm$0.0078&3.7$\pm$1.1&-0.3$\pm$1.1\\
1790.530870&0.643978&0.3674$\pm$0.0056&0.3344$\pm$0.0013&0.9572$\pm$0.0073&6.1$\pm$0.9&0.1$\pm$0.9\\
1791.472450&0.849115&0.3313$\pm$0.0070&0.3277$\pm$0.0014&0.9453$\pm$0.0076&--&--\\
1792.535280&0.080669&0.3808$\pm$0.0084&0.3328$\pm$0.0013&0.9579$\pm$0.0073&--&--\\
\hline
2100.610620&0.199479&0.3737$\pm$0.0071&0.3319$\pm$0.0014&0.9606$\pm$0.0075&--&--\\
2101.635920&0.422856&0.3344$\pm$0.0096&0.3253$\pm$0.0018&0.9335$\pm$0.0110&--&--\\
2102.612400&0.635597&0.3560$\pm$0.0069&0.3281$\pm$0.0015&0.9344$\pm$0.0082&--&--\\
2103.529510&0.835403&0.3419$\pm$0.0054&0.3260$\pm$0.0013&0.9280$\pm$0.0074&--&--\\
2118.614210&0.121830&0.3514$\pm$0.0057&0.3265$\pm$0.0014&0.9339$\pm$0.0081&--&--\\
2124.664630&0.440004&0.3407$\pm$0.0082&0.3278$\pm$0.0017&0.9361$\pm$0.0092&--&--\\
2125.639690&0.652436&0.3459$\pm$0.0051&0.3293$\pm$0.0015&0.9290$\pm$0.0085&--&--\\
2126.631180&0.868447&0.3374$\pm$0.0062&0.3273$\pm$0.0012&0.9252$\pm$0.0069&--&--\\
2127.593570&0.078118&0.3511$\pm$0.0049&0.3288$\pm$0.0012&0.9401$\pm$0.0067&--&--\\
2128.578110&0.292614&0.3490$\pm$0.0058&0.3300$\pm$0.0012&0.9457$\pm$0.0069&--&--\\
2131.572610&0.945011&0.3451$\pm$0.0044&0.3273$\pm$0.0012&0.9412$\pm$0.0070&--&--\\
2132.561810&0.160523&0.3361$\pm$0.0061&0.3270$\pm$0.0013&0.9302$\pm$0.0072&--&--\\
2133.592020&0.384969&0.3523$\pm$0.0055&0.3290$\pm$0.0012&0.9444$\pm$0.0072&--&--\\
2146.566500&0.211654&0.3467$\pm$0.0059&0.3298$\pm$0.0013&0.9401$\pm$0.0075&--&--\\
2147.537680&0.423240&0.3357$\pm$0.0055&0.3283$\pm$0.0013&0.9288$\pm$0.0076&--&--\\
\hline
2482.556520&0.412094&0.3507$\pm$0.0064&0.3285$\pm$0.0011&0.9437$\pm$0.0063&2.8$\pm$0.8&-0.2$\pm$0.8\\
2485.512210&0.056035&0.3385$\pm$0.0083&0.3261$\pm$0.0014&0.9330$\pm$0.0076&12.7$\pm$1.1&-2.4$\pm$1.1\\
2507.629120&0.874534&0.3404$\pm$0.0054&0.3272$\pm$0.0012&0.9341$\pm$0.0066&4.1$\pm$2.5&-3.3$\pm$2.5\\
2509.533370&0.289403&0.3384$\pm$0.0068&0.3263$\pm$0.0016&0.9330$\pm$0.0090&6.2$\pm$0.8&0.5$\pm$0.8\\
2510.636070&0.529643&0.3367$\pm$0.0070&0.3249$\pm$0.0015&0.9287$\pm$0.0083&6.1$\pm$0.9&0.3$\pm$0.9\\
2513.550070&0.164501&0.3520$\pm$0.0064&0.3293$\pm$0.0012&0.9430$\pm$0.0068&10.3$\pm$1.0&0.3$\pm$1.0\\
2514.559260&0.384368&0.3520$\pm$0.0080&0.3288$\pm$0.0014&0.9452$\pm$0.0078&5.4$\pm$0.8&0.6$\pm$0.8\\
2515.562810&0.603007&0.3414$\pm$0.0079&0.3278$\pm$0.0014&0.9384$\pm$0.0079&4.8$\pm$0.8&0.3$\pm$0.8\\
2516.582590&0.825181&0.3515$\pm$0.0062&0.3278$\pm$0.0012&0.9422$\pm$0.0069&3.4$\pm$0.8&-0.1$\pm$0.8\\
2518.601950&0.265129&0.3535$\pm$0.0052&0.3275$\pm$0.0012&0.9367$\pm$0.0067&8.4$\pm$0.8&0.2$\pm$0.8\\
2520.501260&0.678922&0.3498$\pm$0.0060&0.3280$\pm$0.0012&0.9407$\pm$0.0066&2.8$\pm$0.8&0.2$\pm$0.8\\
2523.532240&0.339266&0.3505$\pm$0.0058&0.3293$\pm$0.0012&0.9390$\pm$0.0065&5.0$\pm$0.8&-0.6$\pm$0.8\\
2524.518410&0.554118&0.3548$\pm$0.0056&0.3288$\pm$0.0012&0.9402$\pm$0.0067&7.7$\pm$0.9&-0.7$\pm$0.9\\
2525.463280&0.759972&0.3547$\pm$0.0073&0.3282$\pm$0.0012&0.9402$\pm$0.0068&1.8$\pm$0.8&-0.0$\pm$0.8\\
2526.533220&0.993074&0.3545$\pm$0.0065&0.3281$\pm$0.0012&0.9412$\pm$0.0065&5.9$\pm$0.8&-0.2$\pm$0.8\\

\hline
\footnote{The longitudinal magnetic field measurements are not included for 2012.61 and parts of 2011.67 epochs due to spectropolarimetric errors at NARVAL. See Sect. 3 for details.} 

\end{longtable}
\end{longtab}

\bibliographystyle{aa}
\bibliography{ref}

\begin{thebibliography}{56}
\expandafter\ifx\csname natexlab\endcsname\relax\def\natexlab#1{#1}\fi

\bibitem[{{Auri{\`e}re}(2003)}]{auriere2003}
{Auri{\`e}re}, M. 2003, in EAS Publications Series, Vol.~9, EAS Publications
  Series, ed. J.~{Arnaud} \& N.~{Meunier}, 105

\bibitem[{{Baliunas} {et~al.}(1995){Baliunas}, {Donahue}, {Soon}, {Horne},
  {Frazer}, {Woodard-Eklund}, {Bradford}, {Rao}, {Wilson}, {Zhang}, {Bennett},
  {Briggs}, {Carroll}, {Duncan}, {Figueroa}, {Lanning}, {Misch}, {Mueller},
  {Noyes}, {Poppe}, {Porter}, {Robinson}, {Russell}, {Shelton}, {Soyumer},
  {Vaughan}, \& {Whitney}}]{baliunas1995}
{Baliunas}, S.~L., {Donahue}, R.~A., {Soon}, W.~H., {et~al.} 1995, \apj, 438,
  269

\bibitem[{{Baliunas} {et~al.}(1985){Baliunas}, {Horne}, {Porter}, {Duncan},
  {Frazer}, {Lanning}, {Misch}, {Mueller}, {Noyes}, {Soyumer}, {Vaughan}, \&
  {Woodard}}]{baliunas85}
{Baliunas}, S.~L., {Horne}, J.~H., {Porter}, A., {et~al.} 1985, \apj, 294, 310

\bibitem[{{Barnes}(2007)}]{barnes07}
{Barnes}, S.~A. 2007, \apj, 669, 1167

\bibitem[{{Brandenburg} \& {Subramanian}(2005)}]{brandenburg}
{Brandenburg}, A. \& {Subramanian}, K. 2005, \physrep, 417, 1

\bibitem[{{Brown} {et~al.}(2010){Brown}, {Browning}, {Brun}, {Miesch}, \&
  {Toomre}}]{brown10}
{Brown}, B.~P., {Browning}, M.~K., {Brun}, A.~S., {Miesch}, M.~S., \& {Toomre},
  J. 2010, \apj, 711, 424

\bibitem[{{Brown} {et~al.}(1991){Brown}, {Donati}, {Rees}, \&
  {Semel}}]{brown91}
{Brown}, S.~F., {Donati}, J.-F., {Rees}, D.~E., \& {Semel}, M. 1991, \aap, 250,
  463

\bibitem[{{Catala} {et~al.}(2007){Catala}, {Donati}, {Shkolnik}, {Bohlender},
  \& {Alecian}}]{catala}
{Catala}, C., {Donati}, J.-F., {Shkolnik}, E., {Bohlender}, D., \& {Alecian},
  E. 2007, \mnras, 374, L42

\bibitem[{{Charbonneau}(2010)}]{charbonneau}
{Charbonneau}, P. 2010, Living Reviews in Solar Physics, 7, 3

\bibitem[{{Cincunegui} {et~al.}(2007){Cincunegui}, {D{\'{\i}}az}, \&
  {Mauas}}]{cincunegui}
{Cincunegui}, C., {D{\'{\i}}az}, R.~F., \& {Mauas}, P.~J.~D. 2007, \aap, 469,
  309

\bibitem[{{Donati} \& {Brown}(1997)}]{zdi1997}
{Donati}, J.-F. \& {Brown}, S.~F. 1997, \aap, 326, 1135

\bibitem[{{Donati} {et~al.}(2003){Donati}, {Collier Cameron}, {Semel},
  {Hussain}, {Petit}, {Carter}, {Marsden}, {Mengel}, {L{\'o}pez Ariste},
  {Jeffers}, \& {Rees}}]{donati2003}
{Donati}, J.-F., {Collier Cameron}, A., {Semel}, M., {et~al.} 2003, \mnras,
  345, 1145

\bibitem[{{Donati} {et~al.}(2006){Donati}, {Howarth}, {Jardine}, {Petit},
  {Catala}, {Landstreet}, {Bouret}, {Alecian}, {Barnes}, {Forveille},
  {Paletou}, \& {Manset}}]{donati06}
{Donati}, J.-F., {Howarth}, I.~D., {Jardine}, M.~M., {et~al.} 2006, \mnras,
  370, 629

\bibitem[{{Donati} \& {Landstreet}(2009)}]{donatilandstreet}
{Donati}, J.-F. \& {Landstreet}, J.~D. 2009, \araa, 47, 333

\bibitem[{{Donati} {et~al.}(2000){Donati}, {Mengel}, {Carter}, {Marsden},
  {Collier Cameron}, \& {Wichmann}}]{donati2000}
{Donati}, J.-F., {Mengel}, M., {Carter}, B.~D., {et~al.} 2000, \mnras, 316, 699

\bibitem[{{Donati} {et~al.}(1997){Donati}, {Semel}, {Carter}, {Rees}, \&
  {Collier Cameron}}]{donati1997}
{Donati}, J.-F., {Semel}, M., {Carter}, B.~D., {Rees}, D.~E., \& {Collier
  Cameron}, A. 1997, \mnras, 291, 658

\bibitem[{{Duncan} {et~al.}(1991){Duncan}, {Vaughan}, {Wilson}, {Preston},
  {Frazer}, {Lanning}, {Misch}, {Mueller}, {Soyumer}, {Woodard}, {Baliunas},
  {Noyes}, {Hartmann}, {Porter}, {Zwaan}, {Middelkoop}, {Rutten}, \&
  {Mihalas}}]{duncan90}
{Duncan}, D.~K., {Vaughan}, A.~H., {Wilson}, O.~C., {et~al.} 1991, \apjs, 76,
  383

\bibitem[{{Eisenbeiss} {et~al.}(2013){Eisenbeiss}, {Ammler-von Eiff}, {Roell},
  {Mugrauer}, {Adam}, {Neuh{\"a}user}, {Schmidt}, \& {Bedalov}}]{eisen13}
{Eisenbeiss}, T., {Ammler-von Eiff}, M., {Roell}, T., {et~al.} 2013, \aap, 556,
  A53

\bibitem[{{Ertel} {et~al.}(2012){Ertel}, {Wolf}, {Marshall}, {Eiroa},
  {Augereau}, {Krivov}, {L{\"o}hne}, {Absil}, {Ardila}, {Ar{\'e}valo}, {Bayo},
  {Bryden}, {del Burgo}, {Greaves}, {Kennedy}, {Lebreton}, {Liseau},
  {Maldonado}, {Montesinos}, {Mora}, {Pilbratt}, {Sanz-Forcada}, {Stapelfeldt},
  \& {White}}]{ertel}
{Ertel}, S., {Wolf}, S., {Marshall}, J.~P., {et~al.} 2012, \aap, 541, A148

\bibitem[{{Fares} {et~al.}(2009){Fares}, {Donati}, {Moutou}, {Bohlender},
  {Catala}, {Deleuil}, {Shkolnik}, {Collier Cameron}, {Jardine}, \&
  {Walker}}]{fares}
{Fares}, R., {Donati}, J.-F., {Moutou}, C., {et~al.} 2009, \mnras, 398, 1383

\bibitem[{{Fares} {et~al.}(2012){Fares}, {Donati}, {Moutou}, {Jardine},
  {Cameron}, {Lanza}, {Bohlender}, {Dieters}, {Mart{\'{\i}}nez Fiorenzano},
  {Maggio}, {Pagano}, \& {Shkolnik}}]{hd179949}
{Fares}, R., {Donati}, J.-F., {Moutou}, C., {et~al.} 2012, \mnras, 423, 1006

\bibitem[{{Fares} {et~al.}(2010){Fares}, {Donati}, {Moutou}, {Jardine},
  {Grie{\ss}meier}, {Zarka}, {Shkolnik}, {Bohlender}, {Catala}, \& {Collier
  Cameron}}]{hd189733}
{Fares}, R., {Donati}, J.-F., {Moutou}, C., {et~al.} 2010, \mnras, 406, 409

\bibitem[{{Frasca} {et~al.}(2000){Frasca}, {Freire Ferrero}, {Marilli}, \&
  {Catalano}}]{frasca}
{Frasca}, A., {Freire Ferrero}, R., {Marilli}, E., \& {Catalano}, S. 2000,
  \aap, 364, 179

\bibitem[{{Fuhrmann}(2004)}]{fuhrmann}
{Fuhrmann}, K. 2004, Astronomische Nachrichten, 325, 3

\bibitem[{{Gaidos}(1998)}]{gaidos}
{Gaidos}, E.~J. 1998, \pasp, 110, 1259

\bibitem[{{Gizis} {et~al.}(2002){Gizis}, {Reid}, \& {Hawley}}]{gizis02}
{Gizis}, J.~E., {Reid}, I.~N., \& {Hawley}, S.~L. 2002, \aj, 123, 3356

\bibitem[{{Gomes da Silva} {et~al.}(2013){Gomes da Silva}, {Santos}, {Boisse},
  {Dumusque}, \& {Lovis}}]{gomes}
{Gomes da Silva}, J., {Santos}, N.~C., {Boisse}, I., {Dumusque}, X., \&
  {Lovis}, C. 2013, ArXiv e-prints

\bibitem[{{Hall} {et~al.}(2007){Hall}, {Lockwood}, \& {Skiff}}]{hall07}
{Hall}, J.~C., {Lockwood}, G.~W., \& {Skiff}, B.~A. 2007, \aj, 133, 862

\bibitem[{{Isaacson} \& {Fischer}(2010)}]{isaacson}
{Isaacson}, H. \& {Fischer}, D. 2010, \apj, 725, 875

\bibitem[{{Jeffers} \& {Donati}(2008)}]{sjeffers}
{Jeffers}, S.~V. \& {Donati}, J.-F. 2008, \mnras, 390, 635

\bibitem[{{Kochukhov} {et~al.}(2010){Kochukhov}, {Makaganiuk}, \&
  {Piskunov}}]{kochukhov10}
{Kochukhov}, O., {Makaganiuk}, V., \& {Piskunov}, N. 2010, \aap, 524, A5

\bibitem[{{Leggett} {et~al.}(2008){Leggett}, {Saumon}, {Albert}, {Cushing},
  {Liu}, {Luhman}, {Marley}, {Kirkpatrick}, {Roellig}, \&
  {Allers}}]{browndwarf}
{Leggett}, S.~K., {Saumon}, D., {Albert}, L., {et~al.} 2008, \apj, 682, 1256

\bibitem[{{Livingston} {et~al.}(2007){Livingston}, {Wallace}, {White}, \&
  {Giampapa}}]{livingston}
{Livingston}, W., {Wallace}, L., {White}, O.~R., \& {Giampapa}, M.~S. 2007,
  \apj, 657, 1137

\bibitem[{{L{\'o}pez-Santiago} {et~al.}(2006){L{\'o}pez-Santiago}, {Montes},
  {Crespo-Chac{\'o}n}, \& {Fern{\'a}ndez-Figueroa}}]{lopez}
{L{\'o}pez-Santiago}, J., {Montes}, D., {Crespo-Chac{\'o}n}, I., \&
  {Fern{\'a}ndez-Figueroa}, M.~J. 2006, \apj, 643, 1160

\bibitem[{{Luhman} {et~al.}(2007){Luhman}, {Patten}, {Marengo}, {Schuster},
  {Hora}, {Ellis}, {Stauffer}, {Sonnett}, {Winston}, {Gutermuth}, {Megeath},
  {Backman}, {Henry}, {Werner}, \& {Fazio}}]{luhman}
{Luhman}, K.~L., {Patten}, B.~M., {Marengo}, M., {et~al.} 2007, \apj, 654, 570

\bibitem[{{Marsden} {et~al.}(2014){Marsden}, {Petit}, {Jeffers}, {Morin},
  {Fares}, {Reiners}, {do Nascimento}, {Auri{\`e}re}, {Bouvier}, {Carter},
  {Catala}, {Dintrans}, {Donati}, {Gastine}, {Jardine}, {Konstantinova-Antova},
  {Lanoux}, {Ligni{\`e}res}, {Morgenthaler}, {Ram{\`i}rez-V{\`e}lez},
  {Th{\'e}ado}, {Van Grootel}, \& {BCool Collaboration}}]{marsden}
{Marsden}, S.~C., {Petit}, P., {Jeffers}, S.~V., {et~al.} 2014, \mnras, 444,
  3517

\bibitem[{{Marsden} {et~al.}(2004){Marsden}, {Waite}, {Carter}, \&
  {Donati}}]{marsden2004}
{Marsden}, S.~C., {Waite}, I.~A., {Carter}, B.~D., \& {Donati}, J.-F. 2004,
  Astronomische Nachrichten, 325, 246

\bibitem[{{Messina} \& {Guinan}(2002)}]{photometry}
{Messina}, S. \& {Guinan}, E.~F. 2002, \aap, 393, 225

\bibitem[{{Messina} \& {Guinan}(2003)}]{antisolar}
{Messina}, S. \& {Guinan}, E.~F. 2003, \aap, 409, 1017

\bibitem[{{Meunier} \& {Delfosse}(2009)}]{meunier}
{Meunier}, N. \& {Delfosse}, X. 2009, \aap, 501, 1103

\bibitem[{{Morgenthaler} {et~al.}(2011){Morgenthaler}, {Petit}, {Morin},
  {Auri{\`e}re}, {Dintrans}, {Konstantinova-Antova}, \&
  {Marsden}}]{morgenthaler11}
{Morgenthaler}, A., {Petit}, P., {Morin}, J., {et~al.} 2011, Astronomische
  Nachrichten, 332, 866

\bibitem[{{Morgenthaler} {et~al.}(2012){Morgenthaler}, {Petit}, {Saar},
  {Solanki}, {Morin}, {Marsden}, {Auri{\`e}re}, {Dintrans}, {Fares}, {Gastine},
  {Lanoux}, {Ligni{\`e}res}, {Paletou}, {Ram{\'{\i}}rez V{\'e}lez},
  {Th{\'e}ado}, \& {Van Grootel}}]{morgenthaler12}
{Morgenthaler}, A., {Petit}, P., {Saar}, S., {et~al.} 2012, \aap, 540, A138

\bibitem[{{Parker}(1955)}]{parker}
{Parker}, E.~N. 1955, \apj, 122, 293

\bibitem[{{Petit} {et~al.}(2013){Petit}, {Auri{\`e}re}, {Konstantinova-Antova},
  {Morgenthaler}, {Perrin}, {Roudier}, \& {Donati}}]{2013LNP}
{Petit}, P., {Auri{\`e}re}, M., {Konstantinova-Antova}, R., {et~al.} 2013, in
  Lecture Notes in Physics, Berlin Springer Verlag, Vol. 857, Lecture Notes in
  Physics, Berlin Springer Verlag, ed. J.-P. {Rozelot} \& C.~. {Neiner}, 231

\bibitem[{{Petit} {et~al.}(2009){Petit}, {Dintrans}, {Morgenthaler}, {Van
  Grootel}, {Morin}, {Lanoux}, {Auri{\`e}re}, \&
  {Konstantinova-Antova}}]{petit09}
{Petit}, P., {Dintrans}, B., {Morgenthaler}, A., {et~al.} 2009, \aap, 508, L9

\bibitem[{{Petit} {et~al.}(2008){Petit}, {Dintrans}, {Solanki}, {Donati},
  {Auri{\`e}re}, {Ligni{\`e}res}, {Morin}, {Paletou}, {Ramirez Velez},
  {Catala}, \& {Fares}}]{pascal08}
{Petit}, P., {Dintrans}, B., {Solanki}, S.~K., {et~al.} 2008, \mnras, 388, 80

\bibitem[{{Petit} {et~al.}(2002){Petit}, {Donati}, \& {Collier
  Cameron}}]{petit02}
{Petit}, P., {Donati}, J.-F., \& {Collier Cameron}, A. 2002, \mnras, 334, 374

\bibitem[{{Rees} \& {Semel}(1979)}]{resemel}
{Rees}, D.~E. \& {Semel}, M.~D. 1979, \aap, 74, 1

\bibitem[{{Reiners} \& {Schmitt}(2003)}]{reiners}
{Reiners}, A. \& {Schmitt}, J.~H.~M.~M. 2003, \aap, 398, 647

\bibitem[{{Schr{\"o}der} {et~al.}(2013){Schr{\"o}der}, {Mittag}, {Hempelmann},
  {Gonz{\'a}lez-P{\'e}rez}, \& {Schmitt}}]{schroeder}
{Schr{\"o}der}, K.-P., {Mittag}, M., {Hempelmann}, A.,
  {Gonz{\'a}lez-P{\'e}rez}, J.~N., \& {Schmitt}, J.~H.~M.~M. 2013, \aap, 554,
  A50

\bibitem[{{Semel}(1989)}]{semel89}
{Semel}, M. 1989, \aap, 225, 456

\bibitem[{{Skilling} \& {Bryan}(1984)}]{skilling}
{Skilling}, J. \& {Bryan}, R.~K. 1984, \mnras, 211, 111

\bibitem[{{Valenti} \& {Fischer}(2005)}]{fischer}
{Valenti}, J.~A. \& {Fischer}, D.~A. 2005, VizieR Online Data Catalog, 215,
  90141

\bibitem[{{Waite} {et~al.}(2014){Waite}, {Marsden}, {Carter}, {Petit},
  {Donati}, Jeffers, \& {Boro Saikia}}]{waite}
{Waite}, I., {Marsden}, S.~C., {Carter}, B.~C., {et~al.} 2014, \mnras,
  submitted

\bibitem[{{Wright} {et~al.}(2004){Wright}, {Marcy}, {Butler}, \&
  {Vogt}}]{wright04}
{Wright}, J.~T., {Marcy}, G.~W., {Butler}, R.~P., \& {Vogt}, S.~S. 2004, \apjs,
  152, 261

\bibitem[{{Zuckerman} \& {Song}(2009)}]{age}
{Zuckerman}, B. \& {Song}, I. 2009, \aap, 493, 1149

\end{thebibliography}

\end{document}